\documentclass[aps,preprint, floatfix]{revtex4}

\usepackage{graphicx,color}
\usepackage{psfrag}
\usepackage{amssymb}
\usepackage{eso-pic}
\usepackage[dvips,letterpaper]{geometry} 
\usepackage{amsmath}

\begin{document}

%
%
%
%
\def\oti{{\otimes}}
\def\lb{ \left[ }
\def\rb{ \right]  }
\def\tilde{\widetilde}
\def\bar{\overline}
\def\hat{\widehat}
\def\*{\star}
\def\[{\left[}
\def\]{\right]}
\def\({\left(}		\def\BL{\Bigr(}
\def\){\right)}		\def\BR{\Bigr)}
	\def\BBL{\lb}
	\def\BBR{\rb}
%
%
\def\zb{{\bar{z} }}
\def\zbar{{\bar{z} }}
\def\frac#1#2{{#1 \over #2}}
\def\inv#1{{1 \over #1}}
\def\half{{1 \over 2}}
\def\d{\partial}
\def\der#1{{\partial \over \partial #1}}
\def\dd#1#2{{\partial #1 \over \partial #2}}
\def\vev#1{\langle #1 \rangle}
\def\ket#1{ | #1 \rangle}
\def\rvac{\hbox{$\vert 0\rangle$}}
\def\lvac{\hbox{$\langle 0 \vert $}}
\def\2pi{\hbox{$2\pi i$}}
\def\e#1{{\rm e}^{^{\textstyle #1}}}
\def\grad#1{\,\nabla\!_{{#1}}\,}
\def\dsl{\raise.15ex\hbox{/}\kern-.57em\partial}
\def\Dsl{\,\raise.15ex\hbox{/}\mkern-.13.5mu D}
%
%
\def\ga{\gamma}		\def\Ga{\Gamma}
\def\be{\beta}
\def\al{\alpha}
\def\ep{\epsilon}
\def\vep{\varepsilon}
\def\la{\lambda}	\def\La{\Lambda}
\def\de{\delta}		\def\De{\Delta}
\def\om{\omega}		\def\Om{\Omega}
\def\sig{\sigma}	\def\Sig{\Sigma}
\def\vphi{\varphi}

%
%
\def\CA{{\cal A}}	\def\CB{{\cal B}}	\def\CC{{\cal C}}
\def\CD{{\cal D}}	\def\CE{{\cal E}}	\def\CF{{\cal F}}
\def\CG{{\cal G}}	\def\CH{{\cal H}}	\def\CI{{\cal J}}
\def\CJ{{\cal J}}	\def\CK{{\cal K}}	\def\CL{{\cal L}}
\def\CM{{\cal M}}	\def\CN{{\cal N}}	\def\CO{{\cal O}}
\def\CP{{\cal P}}	\def\CQ{{\cal Q}}	\def\CR{{\cal R}}
\def\CS{{\cal S}}	\def\CT{{\cal T}}	\def\CU{{\cal U}}
\def\CV{{\cal V}}	\def\CW{{\cal W}}	\def\CX{{\cal X}}
\def\CY{{\cal Y}}	\def\CZ{{\cal Z}}

\def\rvac{\hbox{$\vert 0\rangle$}}
\def\lvac{\hbox{$\langle 0 \vert $}}
\def\comm#1#2{ \BBL\ #1\ ,\ #2 \BBR }
\def\2pi{\hbox{$2\pi i$}}
\def\e#1{{\rm e}^{^{\textstyle #1}}}
\def\grad#1{\,\nabla\!_{{#1}}\,}
\def\dsl{\raise.15ex\hbox{/}\kern-.57em\partial}
\def\Dsl{\,\raise.15ex\hbox{/}\mkern-.13.5mu D}
%
%
%
\font\numbers=cmss12
\font\upright=cmu10 scaled\magstep1
\def\stroke{\vrule height8pt width0.4pt depth-0.1pt}
\def\topfleck{\vrule height8pt width0.5pt depth-5.9pt}
\def\botfleck{\vrule height2pt width0.5pt depth0.1pt}
\def\Zmath{\vcenter{\hbox{\numbers\rlap{\rlap{Z}\kern
0.8pt\topfleck}\kern 2.2pt
                   \rlap Z\kern 6pt\botfleck\kern 1pt}}}
\def\Qmath{\vcenter{\hbox{\upright\rlap{\rlap{Q}\kern
                   3.8pt\stroke}\phantom{Q}}}}
\def\Nmath{\vcenter{\hbox{\upright\rlap{I}\kern 1.7pt N}}}
\def\Cmath{\vcenter{\hbox{\upright\rlap{\rlap{C}\kern
                   3.8pt\stroke}\phantom{C}}}}
\def\Rmath{\vcenter{\hbox{\upright\rlap{I}\kern 1.7pt R}}}
\def\Z{\ifmmode\Zmath\else$\Zmath$\fi}
\def\Q{\ifmmode\Qmath\else$\Qmath$\fi}
\def\N{\ifmmode\Nmath\else$\Nmath$\fi}
\def\C{\ifmmode\Cmath\else$\Cmath$\fi}
\def\R{\ifmmode\Rmath\else$\Rmath$\fi}

\def\barray{\begin{eqnarray}}
\def\earray{\end{eqnarray}}
\def\beq{\begin{equation}}
\def\eeq{\end{equation}}

\def\n{\noindent}

\def\Tr{\rm Tr} 
\def\xvec{{\bf x}}
\def\kvec{{\bf k}}
\def\kvecp{{\bf k'}}
\def\omk{\om{\kvec}} 
\def\dk#1{\frac{d\kvec_{#1}}{(2\pi)^d}}
\def\2pid{(2\pi)^d}
\def\ket#1{|#1 \rangle}
\def\bra#1{\langle #1 |}
\def\vol{V}
\def\adag{a^\dagger}
\def\rme{{\rm e}}
\def\Im{{\rm Im}}
\def\pvec{{\bf p}}
\def\fermiS{\CS_F}
\def\cdag{c^\dagger}
\def\adag{a^\dagger}
\def\bdag{b^\dagger}
\def\vvec{{\bf v}}
\def\muhat{{\hat{\mu}}}
\def\vac{|0\rangle}
\def\pcut{{\Lambda_c}}
\def\chidot{\dot{\chi}}
\def\gradvec{\vec{\nabla}}
\def\psitilde{\tilde{\Psi}}
\def\psibar{\bar{\psi}}
\def\psidag{\psi^\dagger} 
\def\m{m_*}
\def\up{\uparrow}
\def\down{\downarrow}
\def\Qo{Q^{0}}
\def\vbar{\bar{v}}
\def\ubar{\bar{u}}
\def\smallhalf{{\textstyle \inv{2}}}
\def\smallsqrt{{\textstyle \inv{\sqrt{2}}}}
\def\rvec{{\bf r}}
\def\avec{{\bf a}}
\def\pivec{{\vec{\pi}}}
\def\svec{\vec{s}} 
\def\phivec{\vec{\phi}}
\def\daggerc{{\dagger_c}}
\def\Gfour{G^{(4)}}
\def\dim#1{\lbrack\!\lbrack #1 \rbrack\! \rbrack }
\def\qhat{{\hat{q}}}
\def\ghat{{\hat{g}}}
\def\nvec{{\vec{n}}}
\def\bull{$\bullet$}
\def\ghato{{\hat{g}_0}}
\def\r{r}
\def\deltaq{\delta_q}
\def\gcharge{g_q}
\def\gspin{g_s}
\def\deltas{\delta_s}
\def\gQC{g_{AF}} 
\def\ghatqc{\ghat_{AF}}
\def\xqc{x_{AF}}
\def\mhat{\hat{m}}
\def\xup{x_2}
\def\xdown{x_1}
\def\sigmavec{\vec{\sigma}}
\def\xopt{x_{\rm opt}}
\def\Lambdac{{\Lambda_c}}
\def\angstrom{{{\scriptstyle \circ} \atop A}     }
\def\AA{\leavevmode\setbox0=\hbox{h}\dimen0=\ht0 \advance\dimen0 by-1ex\rlap{
\raise.67\dimen0\hbox{\char'27}}A}
\def\ratio{\gamma}
\def\Phivec{{\vec{\Phi}}}
\def\singlet{\chi^- \chi^+} 
\def\mhat{{\hat{m}}}

\def\Im{{\rm Im}}
\def\Re{{\rm Re}}

\def\xstar{x_*}

\def\sech{{\rm sech}}

\def\Li{{\rm Li}}

\def\dim#1{{\rm dim}[#1]}

\def\ep{\epsilon}

\def\free{\CF}

\def\Fhat{\digamma}

\def\ftilde{\tilde{f}}

\def\muphys{\mu_{\rm phys}}

\def\xiprime{\tilde{\xi}}

\def\CI{\mathcal{I}}

\title{S-matrix approach to quantum  gases in the unitary limit I: 
 the two-dimensional case}
\author{Pye-Ton How and Andr\'e  LeClair}
\affiliation{Newman Laboratory, Cornell University, Ithaca, NY}

\bigskip\bigskip\bigskip\bigskip

\begin{abstract}

In three spatial dimensions, in the unitary limit of a non-relativistic 
quantum Bose or Fermi gas,
the scattering length diverges.  This occurs at a renormalization
group fixed point,  thus these systems present interesting examples
of interacting scale-invariant models with dynamical exponent $z=2$. 
We study this problem  in two and three spatial dimensions 
using  the S-matrix based approach to the thermodynamics 
we recently developed.  It is well suited to the unitary limit 
where the S-matrix $S=-1$, since it allows an expansion in the inverse
coupling.   We define a meaningful scale-invariant, 
 unitary limit in two spatial dimensions,
where again the scattering length diverges.  In the two-dimensional case, 
the integral equation
for the pseudo-energy becomes transcendentally  algebraic, 
 and we can easily compute
the various universal scaling functions as a function
of $\mu/T$,   such as the energy per particle.     
The ratio of the shear viscosity to the entropy density
$\eta/s$  is above the conjectured lower bound of 
$\hbar/ 4\pi k_B$  for all cases except attractive bosons.   
For  attractive 2-component fermions,  $\eta/s \geq 6.07 \,  \hbar/4\pi k_B$, whereas 
for attractive bosons
$\eta/s \geq  0.4\,  \hbar/4 \pi k_B$.

\end{abstract}

\maketitle

\section{Introduction}

In three spatial dimensions,  in the unitary limit of a quantum Bose or Fermi gas with
point-like interactions,  the
scattering length diverges.    ``Unitary'' here refers to 
the limit on the cross section imposed by  unitarity.  
These systems provide intriguing  examples of interacting, scaling
invariant theories with dynamical exponent $z=2$, i.e. non-relativistic. 
The infinite scattering length occurs at a fixed point of
the renormalization group in the zero temperature theory,
thus the models are quantum critical.   
The only energy scales in the problem are the temperature and
chemical potential,  and  thermodynamic properties 
are expected to reveal universal behavior.   
They can be realized experimentally by tuning the scattering
length to $\pm \infty$ using a Feshbach resonance.
(See for instance \cite{Experiment1,Experiment2} and references
therein.)   They are also thought to occur at the surface of
neutron stars. 

The systems have attracted much theoretical
interest, and remain challenging problems due to
the lack of small parameter for a perturbative expansion,
such as $n^{1/d} a$ or $n^{1/d} r$ where $a$ is the scattering
length, and $r$ the range of the 2-body potential.    
Early works were done by Leggett and Nozi\`eres and 
Schmitt-Rink\cite{Leggett,Nozieres}. 
The universal scaling behavior was studied in \cite{Ho,HoMueller}. 
In 3 dimensions, this is the physics of the BCS/BEC crossover:
since the fixed point occurs for an attractive coupling, 
the fermions may form a bosonic bound state which can subsequently
undergo BEC.    
This cross-over was studied analytically in \cite{Ohashi} 
by considering a model of 2-component fermions coupled to the
bosonic bound state.   
Monte-Carlo studies were performed in 
\cite{Astrakharchik,Bulgac,Burovski,DLee}.
The models can be studied in spatial dimension $2<d<4$\cite{Kolomeisky,Nussinov}
and an epsilon expansion carried out\cite{Nishida,Nikolic}.    
There has also  been some attempts to apply the AdS/CFT correspondence
to these non-relativistic systems\cite{Son,Maldacena,Herzog,Adams}.

In the present work, we describe a new analytic approach to
studying the unitary limit based on our  treatment  of
quantum gases in \cite{PyeTon},  which appears to be  
well suited to the problem since it allows an expansion
in the inverse coupling.   Let us motivate this approach
as follows.   In one spatial dimension, the fixed point occurs
for repulsive interactions.   The model is integrable\cite{Lieb}
and its thermodynamics determined exactly by the so-called
thermodynamic Bethe ansatz (TBA)\cite{YangYang}.     In the TBA,  the free
energy is expressed in terms of a pseudo-energy which is
a solution to an integral equation with a kernel related to
the logarithm of the S-matrix.   In the unitary limit 
the coupling goes to $\infty$ and the S-matrix  $S=-1$.  
The TBA is then identical to a free gas of fermionic particles.   
The formalism developed in \cite{PyeTon} was modeled after
the TBA:    the free energy is expressed as a sum of diagrams
where the vertices represent matrix elements of the logarithm 
of the (zero temperature) S-matrix.   However since  generally 
the N-body S-matrix does not
factorize into 2-body scattering,  the series cannot be summed
exactly as in the TBA.  Nevertheless,  a consistent resummation
of an infinite number of diagrams involving only 2-body scattering,
the so-called foam diagrams,  can serve as a useful approximation
if the gas is not too dense.   The result of summing these diagrams
leads to an integral equation for a pseudo-energy,  as in the TBA;
in fact in 1 spatial dimension the TBA is recovered to lowest order
in the kernel.     Since the formalism is based on the S-matrix, 
it can be very useful for studying the unitary limit where $S=-1$.

In this paper we present the main formulas for the 3-dimensional case, 
however we mainly analyze the 2-dimensional case;   analysis of
the 3d case will be published separately\cite{InPrep}.  
Phase transitions  in two-dimensional Fermi gases were
studied in  e.g. \cite{Petrov2}.
The fixed  point separating the BEC and BCS regimes goes to
zero coupling when $d=2$, thus it is not obvious whether a unitary
limit exists at strong coupling.  As we will argue,  
there is a scale-invariant limit at infinite coupling $g=\pm \infty$
where the S-matrix 
$S=-1$.  This is a meaningful unitary limit at very low ($g=-\infty$)
or very high ($g=+\infty$) energy,  although it does not 
correspond to  a  fixed point in the usual sense of a zero of
the beta function.     The scattering length indeed  diverges
in this limit.   The possibility of this kind of unitary limit
in two dimensions has not been considered before in the literature.

In the next section we describe the unitary limit in 1,2 and 3 dimensions
and its relation to the renormalization group.    In section III, 
we define the interesting scaling functions for the free energy
and single-particle energies by normalizing with respect to
free theories.   In section IV  we describe the unitary limit
of the formalism in \cite{PyeTon} in both two and three dimensions,
where the integral equation becomes scale invariant.   The  $d=2$ 
case is especially simple since the kernel reduces to a constant
and the integral equation becomes algebraic.   Analysis of these
equations in $2d$ is carried out for both infinitely
repulsive or attractive fermions and bosons in sections 
V-VIII.    The extension of our formalism to multiple  species of possibly 
mixed bosonic and fermionic particles is considered in section IX.

Kovtun et. al. conjectured that there is a universal lower bound
to the ratio of the shear viscosity to the entropy density,
\beq
\label{boundzero}
\eta/s \geq \frac{\hbar}{4 \pi k_B}
\eeq
 where $k_B$ is Boltzmann's
constant\cite{Kovtun}.    This was based on 
the AdS/CFT correspondence for {\it relativistic} theories in
3 spatial dimensions,
and the bound is saturated for certain supersymmetric gauge theories. 
Counterexamples to the $\eta/s$ bound were suggested to be non-relativistic\cite{Kovtun},
however no known fluid violates the bound.    It thus interesting to study this ratio for non-relativistic theories,
and in particular for $2d$ theories where no conjecture have been put
forward. 
  It has also been suggested that
the gases in the unitary limit may represent the most perfect fluid,  i.e.
with the lowest value of $\eta/s$.   
  For the spacetimes considered thus far for a non-relativistic
AdS/CFT correspondence in 2d,   the result found is that $\eta/s$ is
exactly $1/4\pi$\cite{Herzog,Adams}.   We analyze $\eta/s$ for the attractive fermionic case 
as a function
of $\mu /T$ in section VI.    If one disregards  unphysical,  potentially metastable regions,
one finds  $\eta/s \geq  6.07 \,  \hbar/ 4 \pi k_B$. 
The other cases of repulsive bosons or fermions also satisfy the bound. 
However the  attractive boson   is below it:  
  $\eta/s \geq 0.4\, \hbar / 4\pi k_B$.

\section{Renormalization group and the unitary limit}

The models considered in this paper are the simplest 
models of non-relativistic bosons and fermions with quartic
interactions.    The bosonic model is defined by the action
for a complex scalar field $\phi$. 
\beq
\label{bosonaction}
S =  \int d^d \xvec dt \(  i \phi^\dagger  \d_t \phi - 
\frac{ |\vec{\nabla} \phi |^2}{2m}  - \frac{g}{4} (\phi^\dagger  \phi)^2 \)
\eeq
(Throughout this paper, $d$ refers to the number of spatial dimensions). 
For fermions, due to the fermionic statistics, 
one needs at least a 2-component field 
$\psi_{\up , \down} $:
\beq
\label{fermionaction}
S = \int d^d \xvec dt \(  \sum_{\alpha=\up, \down}  
i \psi^\dagger_\alpha \d_t  \psi_\alpha  - 
\frac{|\vec{\nabla}  \psi_\alpha|^2}{2m}   - \frac{g}{2} 
\psi^\dagger_\up \psi_\up \psi^\dagger_\down \psi_\down \) 
\eeq
In both cases,  positive $g$ corresponds to repulsive interactions.   

The bosonic theory only has a $U(1)$ symmetry.   The fermionic theory
on the other hand has the much larger SO(5) symmetry.   
This is evident from the work\cite{Kapit} which considered a relativistic
version, since  the same arguments apply to a non-relativistic kinetic term. 
This is also clear from the work\cite{Nikolic} which considered 
an $N$-component version with Sp(2N) symmetry,  and noting that
Sp(4) = SO(5).

The free versions of the above theories have a 
scale invariance with dynamical exponent $z=2$, 
i.e. are invariant under
 $$t\to \Lambda^{-2} t, ~
\xvec \to \Lambda^{-1} \xvec$$ 
  At a renormalization 
group fixed point, i.e. quantum critical point, 
the model is expected to have
the same scale invariance.  It is natural to 
define scaling dimensions $\dim X$ in units of 
inverse length or wave-vector $\kvec$, 
i.e. $\dim{\kvec} = 1$,  $\dim{\xvec} = -1$, and
$\dim{t} = -2$.   Requiring the action to have
zero scaling dimension gives
$\dim{\phi} = \dim{\psi} = d/2$ and 
$\dim{g} = 2-d$.    The interaction is thus
relevant for $d<2$.

\def\ghat{\hat{g}}

\subsection{3d case}

The renormalization group behavior can be inferred from the
coupling constant dependence of the S-matrix, or equivalently
the kernels in section IV.   For completeness we compute
the beta function for arbitary $d$ in Appendix A using
conventional  methods.    
  Consider first a single boson in  $d=3$ dimensions.
The kernel in eq. (\ref{G3d}) depends on the renormalized
coupling $g_R$ given in eq. (\ref{gR}), where $\Lambda$ is
a high momentum cutoff.   
The kernel $G$ is related to the logarithm of the 2-body S-matrix
as in eq. (\ref{G2_def}), and from this we can deduce the
S-matrix function:
\beq
\label{3dS}
S(|\kvec - \kvec'|) =   \frac{16 \pi/m g_R  - i  |\kvec - \kvec'|}
{16\pi/ m g_R  + i  |\kvec - \kvec'|}
\eeq
Galilean invariance, for equal mass particles, 
 implies $S$ is only a function 
of the relative momenta $\kvec, \kvec'$ of the two in-coming particles.   
Unitarity of the S-matrix amounts to $S^* S = 1$.  
   Defining 
$g= \hat{g} \Lambda^{2-d}$, and requiring $g_R$ to be independent
of $\Lambda$ gives the beta-function:
\beq
\label{betafun}
\frac{ d \ghat}{d \ell}  =  - \ghat - \frac{m}{4\pi^2} \ghat^2 
\eeq
where $\ell =  - \log \Lambda$ is the logarithm of a length scale. 
The above result agrees with the  calculation in Appendix A. 
With the normalization of $g$ in the fermionic case 
as given in  (\ref{fermionaction}),  the resulting  beta function
 is the same as above. 
 
\begin{figure}[htb] 
\begin{center}
\hspace{-15mm}
\psfrag{B}{$g_*$} 
\includegraphics[width=10cm]{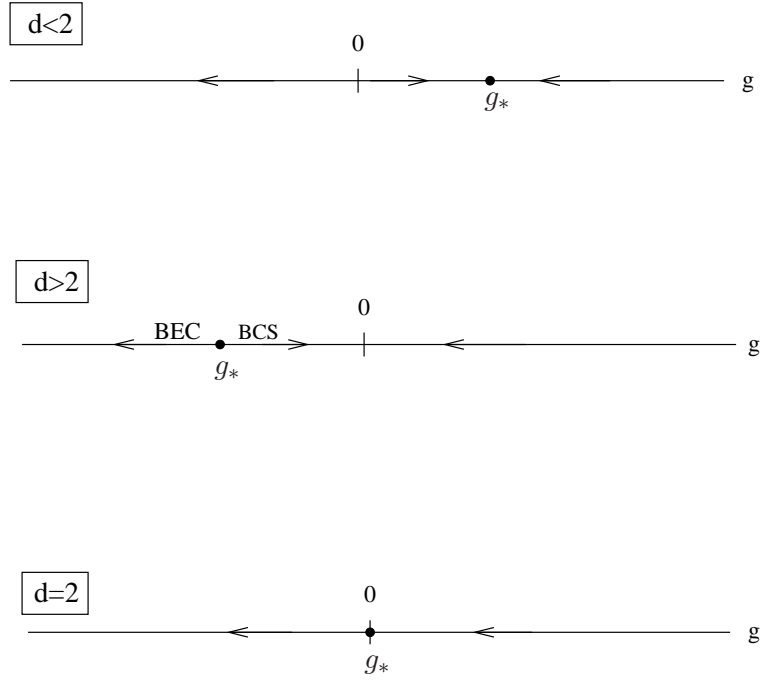} 
\end{center}
\caption{Renormalization group flows;  arrows indicate the flow
to low energy.}  
\vspace{-2mm}
\label{RG} 
\end{figure}

As discussed in \cite{Kolomeisky,Nikolic}, the  beta-function in $d$ dimensions
(see Appendix A) has
the following implications.   See Figure \ref{RG}.   For $d<2$,  there is an infra-red
stable fixed point at the positive value  
\beq
\label{ghatstar}
\ghat_* = (2-d) \pi^{d/2} \Gamma (d/2) 2^d / m
\eeq
corresponding to repulsive interactions.   When $d=1$,  this 
fixed point is approached from $g = \pm \infty$ when the cutoff 
$\Lambda$ is infinite.   In the limit of $g\to \pm \infty$ the 
S-matrix, as given in
eq. (\ref{1dS}) becomes 
 $S=-1$,  which is the unitary limit.   In the 1-dimensional
thermodynamic Bethe ansatz,  the thermodynamics is simply that
of a free, one-component, fermionic gas, even though
the original particle was a boson.    

For $d>2$,  there is an {\it ultra-violet} stable  fixed point
at the {\it negative} coupling $\ghat_*$,
 i.e. an attractive 
interaction.     A bosonic gas
with attractive interactions  can be  unstable.
However for a fermionic gas,  the attractive interaction
can be balanced by the Fermi pressure, and the gas can be stable
against  collapse.  

We now describe the implications of the above fixed point on
the scattering length.    Consider a single boson.  
A  straightforward calculation 
of the  differential cross section in $d$ spatial dimensions gives
\beq
\label{cross}
\frac{d \sigma}{d\Omega} =   \frac{ m^2  k^{d-3}}{4 (2\pi)^{d-1} }  
| \CM (k) |^2 
\eeq
where $\CM (|\kvec  - \kvec' |)$ is the scattering amplitude.
In the above formula $k$ is  the momentum of one of the particles in
the center of mass frame, i.e. $2k = |\kvec - \kvec'|$.  
The $\CM$ are an important ingredient of the kernels of the integral
equations in section IV,  and were computed to all orders in
\cite{PyeTon} for instance.  In $d=3$,   we equate $\sigma = \pi a^2$.
   This gives
\beq
\label{a3d}
a (k) =  \frac{m}{2\pi}   \frac{ g_R}{\sqrt{ 1 + (m g_R k /8\pi)^2 }}
\eeq
where 
$g_R$ is the renormalized coupling (\ref{gR}).  
If $a(k)$ is measured at very small momentum transfer $|\kvec - \kvec'|
\approx 0$,  this leads
to the definition of the scattering length 
\beq
\label{a3d.2}
a =  \frac{m g_R}{2\pi}  =   \frac{mg}{2\pi (1 + mg\Lambda/4\pi^2)} 
\eeq
where $\Lambda$ is the ultra-violet cutoff.   
In order for $a$ to diverge,  $g$ must be negative.   In fact,
at precisely the fixed point $g_* = \ghat_*/\Lambda$,  
$a\to \pm \infty$, depending on from which side $g_*$ is approached.
When $g=g_*^-$, i.e. just less than $g_*$,  then $a\to \infty$, 
whereas when $g=g_*^+$, $a \to -\infty$. 
 The case $g_R = -\infty$ is on the BCS side of the crossover,
whereas $g_R = + \infty$ is on the BEC side.
See Figure \ref{RG}.

The S-matrix (\ref{3dS}) has a pole at $k=16 \pi i/mg_R$. 
  Since physical bound states correspond to 
poles at $\Im ( k ) >0$, 
 the bound state exists only on the 
BEC side of the critical point,
with energy
\beq
\label{Ebound}
E_{\rm bound - state} =  -  \frac{128 \pi^2}{m^3 g_R^2}
\eeq
In the BEC to BCS crossover 
from $1/a = 0^+$ to $1/a =  1/0^-$ the binding energy goes
to zero and the bound state disappears.
Nevertheless, the crossover is expected to be smooth, if   
on the BEC side one  includes  the bound state in
the thermodynamics.

\subsection{2d case}

The $d=2$ case is somewhat more subtle due to the marginality of 
the coupling $g$. 
The fixed point $\ghat_* =0$ and the RG flows are depicted in
Figure \ref{RG}.    The fixed point at $\ghat_* =0$ is just the
free field theory,  thus there is no analog of the 
BEC/BCS cross-over.   One can nevertheless formally define
the unitary limit as $S=-1$, as in $1d$ and $3d$.   
In this subsection we explore this possibility and 
interpret it using the renormalization group. 
As we'll see,  this limit occurs at $g=\pm \infty$,
and the scattering length diverges.

First begin with the beta-function in 2d, eq. (\ref{A.4}):
\beq
\label{2dfp.1}
\frac{ dg }{d \log \Lambda} = \frac{m g^2}{4\pi}
\eeq
Let $g=g_0$ at some arbitrary scale $\Lambda_0$.  Integrating
the beta function one finds:
\beq
\label{2dfp.2}
g(\Lambda)  =  \frac{  g_0}{1 - \frac{m g_0}{4\pi} \log (\Lambda/ \Lambda_0 )}
\eeq
Note that $g$ diverges at the scale:
\beq
\label{lambdastar}
\Lambda_* =   \Lambda_0 e^{4\pi/mg_0}
\eeq
This is the familiar Landau pole.   Whereas in for example the
relativistic 
$\phi^4$ theory in $d=3$ where the Landau pole is unphysical
due to higher order corrections,  here the beta function 
(\ref{2dfp.1}) is exact\cite{Nikolic},  thus this divergence is physical.  
There are two cases to consider: 

\bigskip
\n {\it Attractive case:}  ~  If $g_0$ is negative,  
$\Lambda_* < \Lambda_0$,  and thus $g=-\infty$ occurs
in the infra-red. 

\bigskip

\n {\it Repulsive case:} ~  If $g_0$ is positive,  $\Lambda_* > \Lambda_0$,
and $g= +\infty$ occurs in the ultra-violet.  

\bigskip
\n Both cases are consistent with the flows depicted in Figure \ref{RG}. 

In fact,  the scale $\Lambda_* =  \Lambda e^{4\pi/mg}$ is an 
RG invariant:
\beq
\label{2dfp.3}
\frac{ d \Lambda_*}{d \log \Lambda} = 0
\eeq
and is the natural coupling constant in this problem that
physical quantities are expressed in terms of.    
The S-matrix can be inferred from the kernel in eq. (\ref{kernel.2d}):
\barray
\nonumber 
S( |\kvec| ) &=& \frac{  4\pi /mg  + 
\log( 2 \Lambda/|\kvec|  ) - i\pi/2}
{   4\pi /mg  + \log( 2 \Lambda/|\kvec| ) + i\pi/2}
\\
&~&
\label{2dS}
\\
&=& \frac{  \log (2 \Lambda_*/|\kvec| ) - i \pi/2}{\log
 (2 \Lambda_*/|\kvec| ) + i \pi/2}
\earray 
where $|\kvec| = |\kvec_1 - \kvec_2 |$ is the relative momentum
of the two incoming particles.
Near the fixed point $g=0^-$,  $\Lambda_* = 0$ 
($\Lambda_* = \infty$)  and the
S-matrix $S=1$, consistent with the free theory.  
The same applies to the other side of the fixed point $g=0^+$ 
where $\Lambda_* = \infty$.    On 
the other hand,  consider the low energy limit of the
attractive case where $g=-\infty$.  Here $\Lambda_* = \Lambda$,
and this scale is effectively an infra-red cutoff.  
Thus at low energies  $|\kvec_1 - \kvec_2| \approx 2 \Lambda_*$ and $S=-1$.  
Similar arguments apply to the high energy limit of
the  repulsive  case $g=+\infty$.   It is clear that, unlike in $3d$,
 this definition
of the unitary limit does not correspond to a renormalization group
fixed point in the usual sense of a zero of the beta function
and is somewhat engineered; nevertheless
it defines a scale-invariant theory.

Let us turn now to the scattering length.   Using the scattering
amplitude $\CM$ computed in \cite{PyeTon}, the cross section 
for relative momenta $ 2k $ is:
\barray
\label{2dunit.2}
\sigma (k) &=&  \frac{m^2 g^2}{8k} \, 
\inv{ ( 1 + \frac{mg}{4\pi} \log (\Lambda/k))^2 + \pi^2 } 
\\ 
&=& \frac{2 \pi^2}{k} 
\inv{ \log^2 (\Lambda_* / k)  +  16\pi^4/m^2 g^2}
\earray
Near the scale $k=\Lambda_*$:
\beq
\label{sigstar}
\sigma (\Lambda_* ) = \frac{2}{\Lambda_*} \( \frac{ mg}{4\pi} \)^2 
\eeq
Equating $\sigma$ with a scattering length $a$, as appropriate
to $2d$,  one sees that $a$ diverges in the unitary limit
$g\to \pm \infty$.   

Because of the logarithmic dependence on $|\kvec_1 - \kvec_2|$,
the S-matrix does not have a pole.  However the denominator
is zero when $|\kvec_1 - \kvec_2| = 2k = 2 i \Lambda_*$, which
is a remanant of the bound state pole in $3d$.    The energy of
this quasi-bound state is $-2 \Lambda_*^2/m$.  Note that this quasi-bound
state disappears as $g_0 \to 0^-$, since in this limit
$\Lambda_* \to 0$,  and this is analagous to the situation
in $3d$.    We thus  expect that the ``attractive'' case of the unitary limit
should be better behaved since this bound state disappears,  in contrast to
the repulsive case where $\Lambda_* \to \infty$ as $g_0 \to 0^+$,
and this will be born out of our subsequent analysis.

\section{Thermodynamic  scaling functions  at the quantum critical point}

At a quantum critical point, the only length scales of
the quantum gas are the thermal wavelength 
$\lambda_T =  \sqrt{2\pi/mT}$  and the length scale $n^{1/d}$ set
by the density $n$.    Equivalently,  one can express 
physical properties in terms of the only two energy scales,
the temperature $T$ and the chemical potential $\mu$,  since
the density is a function of $T,\mu$.   

In order to fix normalizations in a meaningful way,   it
is useful to consider the simplest theories with $z=2$ scale
invariance:  free, non-relativistic bosons and fermions. 
We set $k_B = \hbar =1$, except in a few select formulas.    
The free energy density is given by the well-known expression
for each particle:
\beq
\CF =  \frac{s}{\beta} \int  \frac{d^d \kvec}{(2\pi )^d} \, 
\log \( 1- s e^{-\beta (\omega_\kvec - \mu )} \) 
\eeq
where $\beta = 1/T$, $\omega_\kvec = \kvec^2 / 2m $ 
 and $s=1,-1$ corresponds to bosons, fermions
respectively.   
The integrals over wave-vectors can be expressed in terms of
polylogarithms $\Li_\nu (z)$, where $z=e^{\beta \mu}$ is a fugacity, 
using $\int d^d \kvec = 2 \pi^{d/2} /\Gamma(d/2) \int dk  k^{d-1} $
and the integrals
\barray
\int_0^\infty dx ~  \frac{z x^{\nu -1} }{e^x -z } &=& \Gamma (\nu) \Li_\nu (z) 
\nonumber \\
\label{LiInt}
\int_0^\infty dx  ~  \frac{z x^{\nu -1} }{e^x +z } &=& -\Gamma (\nu) \Li_\nu (-z)
\earray
valid for $\Re (\nu) > 0$.   
The result, as expected, is proportional to $T/\lambda_T^d$:
\beq
\CF =  - s T \( \frac{mT}{2\pi} \)^{d/2}  \Li_{(d+2)/2} (s z )
\eeq

There are two important limits to consider.   For Bose gases near
Bose-Einstein condensation,  physically the interesting limit is
$\mu / T  \to 0$.  Since $\Li_\nu (1) = \zeta (\nu)$, 
where $\zeta$ is Riemann's zeta function, 
this leads us to define the scaling functions 
$c_d (\mu / T )$:
\barray
\CF  &=&  - \frac{\pi m T^2}{12}    ~  c_2 (\mu /T) ~~~~~~~~~~~~~~~~~~~(d=2) 
\label{freec}
\\
\nonumber
\CF  &=&  -  \frac{\zeta (5/2) m^{3/2} T^{5/2} }{(2\pi)^{3/2}}  
~ c_3 (\mu /T) 
~~~~~~~(d=3)
\earray
where we have used $\zeta (2) = \pi^2/6$.   
With the above normalizations,  $c_d =1$ for a single 
free boson when $\mu /T =0$.

The above formulas are well-defined for fermions at zero chemical potential.
Using 
\beq
\label{Lifermion}
- \Li_\nu (-1) =  \( 1 - \inv{2^{\nu -1}} \) \zeta (\nu)    
\eeq
one finds as $\mu/T \to 0$:
\beq
\label{cfermion} 
c_2 =  \inv{2} ,  ~~~~~~ c_3 =  1 - \inv{2 \sqrt{2}}  ~~~~~~~~~~
( {\rm free ~fermions} )
\eeq
It should be pointed out that the coefficients $c_d$ are analogous
to the Virasoro central charge for relativistic systems in $d=1$,
as discussed in \cite{Leclair}.

The other interesting limit is $T/\mu \to 0$, i.e. $z\to \infty$ 
or $z\to 0$,  depending on the sign of the chemical potential.
  Here the
scaling forms are naturally based on the zero temperature 
degenerate free fermion gas,  where $\mu>0$ is the Fermi energy. 
In fact, the function  $\Li_\nu (z)$  has a branch cut along the
real axis from $z=1$ to $\infty$ and the bosonic free energy is
ill-defined at $z=\infty$, in contrast with fermions.       
Using the analytic continuation of the asymptotic behavior
\beq
\label{largez}
- \Li_\nu (-z) \approx  \frac{\log^\nu (z)}{\Gamma(\nu +1) } ~~~~~~ 
{\rm as } ~ z \to \infty 
\eeq
from positive integer $\nu$ to half-integer values,  
we define the scaling functions $b_d$ as follows:
\barray
\CF &=& - \frac{m\mu^2 }{4\pi}   ~   b_2 (T/\mu)~~~~~~~~~~~~~~~~~~~~(d=2)
\label{tzero}
\\  \nonumber 
\CF &=&  - \frac{2 \sqrt{2} m^{3/2} \mu^{5/2}  }{15 \pi^2 }  ~ 
 b_3 (T/ \mu ) ~~~~~~~~~(d=3) 
\earray
The above normalizations are defined such that $b_d (0) = 1$ for
a single free fermion.
One can  verify  that the above $d=3$ expression  is the standard 
result for a zero temperature, single component, degenerate fermion gas 
where $\mu$ is the Fermi energy.

Other thermodynamic quantities follow from the free energy.
The pressure $p = - \CF$.   The density is 
$n = - \d \CF / \d \mu $  and the entropy density 
is $s = - \d \CF / \d T$.     Using the scaling form in eq.
(\ref{tzero}), one obtains for $d=3$:
\barray
\label{density3d}
n &=&  \frac{2 \sqrt{2}}{15 \pi^2}  (m \mu )^{3/2} \(  \frac{5}{2} b_3 -
\frac{T}{\mu} b_3' \)   
\\
\nonumber 
s &=&  \frac{2 \sqrt{2}}{15 \pi^2}  (m \mu)^{3/2} \,  b_3' (T/\mu ) 
\earray
where $b'$ is the derivative with respect to its argument $T/\mu $.
(Henceforth, $g'$ will always refer to the derivative  
of the function $g$ with respect to it's argument $\mu/T$ or 
$T/\mu$ as defined above.) 
The analogous formulas in 2 dimensions are:
\barray
\label{density2d}
n &=&  \frac{m \mu}{4\pi} \(  2  b_2 -  \frac{T}{\mu}  b_2' \) 
\\ \nonumber
s  &=&   \frac{ m \mu}{4\pi} \,  b_2' 
\earray

The energy density $\epsilon$  follows from the relation
$\epsilon = T s + \mu n  + \CF $:
\beq
\label{ener}
\epsilon = - \frac{d}{2} \CF 
\eeq
It is interesting to note that the above result,  which in terms of
the pressure is simply $\epsilon = p d/2$,  is  
a consequence of the
mechanics of {\it free}  gases.    This shows  that 
this relation continues to hold for interacting gases at
a quantum critical point, as pointed out by Ho\cite{Ho}.

Also of interest is the energy per particle $\ep_1 = \ep/n$:
\beq
\label{enpart.1}
\ep_1 =  - \frac{d}{2}  \frac{\CF}{n} 
\eeq
Consider first the limit $T/\mu \to 0$, with $\mu$ positive. 
 In this limit,  if $b$ is a smooth function of $T/\mu$
as $T/\mu \to 0^+$,  then the $b'$ 
terms in the density vanish and one simply obtains
\beq
\label{enpart.2}
\lim_{T / \mu \to 0^+} \ep_1 =  \frac{d}{d+2} \mu 
\eeq   
which is the same result as for a free gas, where for fermions
$\mu$ is equal to the Fermi energy $\ep_F$. 
The above formula is usually not appropriate to the
$T\to 0$ limit when $\mu$ is negative.   For the interacting
gas $\mu \neq \ep_F$, so this leads us to define the scaling 
functions 
$\xi$:
\beq
\label{enpart.3}
\xi_d (T/ \mu)  =  \frac{d+2}{d} \frac{\ep_1}{\ep_F} 
\eeq
As $T / \mu \to 0^+$,  the functions $\xi_d$ should become 
universal constants, and for free fermions $\xi_d (0^+) = 1$.    
The Fermi energy $\ep_F$ can be defined based on its relation 
to density in  the
zero temperature free fermion gas.  
 For bosons,  one can formally use the same definition.  
For $d=2$,  $\ep_F = 2 \pi n/m$,  whereas for
$d=3$, $\ep_F = (3 \pi^2 n/\sqrt{2})^{2/3}/m$.  
This leads to the definitions:  
\barray
\label{xidefs} 
\xi_2  &=&  \( \frac{m\mu}{2\pi n} \)^2 ~b_2 (T/ \mu) 
\\ 
\nonumber
\xi_3 &=&  \( \frac{\sqrt{2}}{3 \pi^2 n} \)^{5/3} ( m \mu)^{5/2} ~b_3 (T/\mu)
\earray

Next consider the energy per particle in the
limit $\mu/ T \to 0$.   Here it is more
appropriate to use the form in eq. (\ref{freec}), which gives
\beq
\label{enpart.3b}
\ep_1 =  \frac{d}{2} \frac{c_d}{c'_d} \, T 
\eeq
The expression  for $\ep_1$ for free fermions in the
limit $\mu /T \to 0$  in $d=3$ leads 
us now to define $\xiprime$:
\beq
\label{enpart.4}
\ep_1 =  \frac{3}{2} 
\( \frac{2\sqrt{2} -1}{2 \sqrt{2} -2} \)
\frac{ \zeta (5/2)}{ \zeta (3/2) } \, T  ~ \xiprime_3 (\mu/T) 
\eeq
With this normalization, for free fermions, as $\mu / T \to 0$, 
$\xiprime_3 = 1$.

In two dimensions, $\ep_1 $ goes to zero  for free bosons as $\mu/ T \to 0$ 
since it is proportional to $1/\zeta (1)$.   However 
$\ep_1$ is finite for fermions in this limit.  
Using $\lim_{d\to 2}  (2^{d/2} -2) \zeta (d/2) = 2 \log 2$, 
we define
\beq
\label{enpart.5}
\ep_1 =     \frac{\pi^2 T}{12 \log 2}  
 ~\xiprime_2 (\mu /T) ~~~~~~~~~~~(d=2)
\eeq
With the above normalization,  $\xiprime_2 (0) = 1$ for free fermions.

Henceforth we drop the subscripts $2,3$ 
indicating  the spatial dimension on the functions $c_2 , c_3$ etc., 
since in the sequel we will carry out the analysis of the $d=2$ case only.

\section{Thermodynamics from the S-matrix in the unitary limit for d=2,3}

In the formalism developed in \cite{PyeTon},  the contributions
to the free energy can be expressed as vacuum diagrams where
the vertices are  matrix elements of the logarithm of the
S-matrix,  and the lines are the filling fractions.   
There are vertices with $2N$  lines representing $N$-body
scattering for any $N$.    
The main result obtained in \cite{PyeTon}, and reviewed below,
 is the formula for
the free energy density in 
 a self-consistent 
resummation of the 2-body scattering ``foam diagrams''. 
The 2-body scattering
can be computed to all orders in the coupling in a standard
zero-temperature calculation.     
Clearly this result  is still an approximation in that 
higher N-body scattering, and 2-body scattering
contributions not of the foam diagram type,  are neglected.
Nevertheless,    this approximation 
is expected to be valid when the gas is sufficiently dilute.

The main ingredients of the formalism are as follows.   
Define the filling fractions as a function of a
pseudo-energy $\vep (\kvec )$ 
\beq
\label{fill}
f(\kvec )  =  \inv{ e^{\beta \vep (\kvec ) } -s }
\eeq
which determine the density:
\beq
\label{dens}
n = \int \frac{d^d \kvec}{(2\pi)^d} ~ \inv{ e^{\beta \vep  (\kvec ) } -s }
\eeq
The consistent summation of 2-body scattering leads to 
an integral equation for the 
 pseudo-energy $\vep (\kvec)$:
\beq
\label{pseudo.energy}
\vep ({\kvec})  = \omega_{\kvec} - \mu -\frac{1}{\beta} \log \left ( 1 +
\beta  \int(d\kvec')  G(\kvec, \kvec')   
\frac{ e^{\beta (\vep (\kvec') - \omega_{\kvec'}+\mu )}}
{e^{\beta \vep  (\kvec' ) } -s } \right )
\eeq
The kernel $G$ in this equation  is related to the logarithm of the 2-body 
S-matrix $\hat{S}$ as follows:
\begin{equation}
2 \pi \delta \left (E - \frac{1}{2m}(\kvec^2 + \kvec'^2) \right ) V
 \; G (\kvec,\kvec') \equiv  -i \langle \kvec, \kvec' \vert 
 \log\hat{S}(E) \vert \kvec, \kvec' \rangle.
\label{G2_def}
\end{equation}
where $V$ is the spatial volume and $(d\kvec) =  d^d \kvec /  (2\pi)^d$. 
The kernel has the following structure:
\beq
\label{Gstructure}
G  =  - \frac{i}{\CI} \log ( 1 + i \CI \CM)
\eeq
where $\CM$ is the scattering amplitude and $\CI$ represents
the available phase space for two-body scattering.   The argument
of the $\log$ can be identified as the S-matrix function. 
Finally, the  free energy density  then takes the simple form:
\beq
\label{freefoam}
\free 
\nonumber
=  - \inv{\beta}  \int (d\kvec) 
\(  s \log ( 1 + sf)  - \inv{2}  \( \frac{f-f_0}{1+ s f_0} \)  \)
\eeq
where $f_0$ is the filling fraction of the free theory:
\beq
\label{fzero}
f_0 (\kvec ) =  \inv{e^{\beta ( \omega_\kvec-\mu )} - s} 
\eeq

Consider first the 1-dimensional bosonic case.  The model is integrable,
which implies that the N-body S-matrix factorizes into 2-body S-matrices,
and the exact free energy is given by the 
thermodynamic Bethe ansatz\cite{YangYang}.   The 2-body S-matrix is 
\beq
\label{1dS}
S(k - k') =  \frac{ k-k' -ig/4}{k-k' + ig/4}
\eeq
In the unitary limit $g\to \pm \infty$,  $S=-1$,  and the  
thermodynamic Bethe ansatz reduces to that of a free gas of fermionic
particles.

In 3 spatial dimensions the exact kernel for a single component 
boson is the following\cite{PyeTon}:
\beq
\label{G3d}
\begin{split}
G (|\kvec -\kvec'| ) &= -  \frac{8 \pi i }{m \vert \kvec - \kvec' \vert} \log 
\left(
\frac{1 - \frac{i m g_R}{16 \pi} \vert \kvec - \kvec' \vert}{1 + 
\frac{i m g_R}{16 \pi} \vert \kvec  - \kvec'  \vert}\right) \\
&= - \frac{16 \pi}{m \vert \kvec  - \kvec'  \vert} \arctan \left
 (\frac{m g_R}{16 \pi} \vert \kvec  - \kvec' \vert \right)
\end{split}
\eeq
where $g_R$ is a renormalized coupling:
\beq
\label{gR}
\frac{1}{g_R} = \frac{1}{g} + \frac{m \Lambda}{4 \pi^2}.
\eeq
with $\Lambda$ an ultra-violet momentum cutoff.   
As in the 1-dimensional case, as discussed in section II, 
 the unitary limit corresponds to 
$g_R \to \pm \infty$ where $S=-1$.    Thus in the unitary limit the
kernel becomes
\beq
\label{unitaryG3d}
G  (\kvec , \kvec') =  \mp  \frac{8 \pi^2}{m |\kvec - \kvec' |}
~~~~~~~~~~(d=3)
\eeq
 where $-,+$   corresponds to $g$ being  just below, above
$g_*$, where  the scattering length $a=+\infty$ (BEC side) 
and $a=-\infty$ (BCS side)
respectively.   It should be kept in mind that the underlying
interactions are attractive in both cases since the fixed point occurs at
negative $g$.

 In two spatial dimensions the single boson 
kernel obtained in \cite{PyeTon} is 
\begin{equation}
\begin{split}
G (|\kvec| ) &=
- \frac{4 i}{m} \log \left ( \frac{1 + 
	\frac{mg}{4 \pi} \left (
		\log \left ( \frac{2 \Lambda}{\vert \kvec \vert
} \right ) - i \pi/2
	\right ) }
	{1  +
	\frac{mg}{4 \pi} \left (
		\log \left ( \frac{2 \Lambda}{\vert \kvec  \vert}
 \right ) + i \pi /2 
		\right )}  \right ) \\
& =  - \dfrac{8}{m} \arctan \left( 
	\frac{mg/8}
		{1 
		+ \frac{mg}{4\pi} 
			\log \left( \frac{2 \Lambda}{\vert \kvec 
 \vert} \right)
		}
	\right) 
\\
& = - \frac{8}{m}  \arctan  \(  \frac{2 \pi}{\log
 (2 \Lambda_* /|\kvec|)} \)
\end{split}
\label{kernel.2d}
\end{equation}
where $\Lambda_*$ is defined in eq. (\ref{lambdastar}),
and $|\kvec| = |\kvec_1 - \kvec_2| $  is the relative momentum. 
In the unitary limit $g\to \pm \infty$,  the theory is at the scale
$\Lambda_*$ and one should
 consider $|\kvec_1 - \kvec_2| \approx 2 \Lambda_*$.  
 The result is that $G$ becomes a constant in this unitary limit: 
\beq
\label{unitaryG2d}
G  (|\kvec|  )  =  \mp  \frac{4\pi}{m} 
~~~~~~~~~~(d=2) 
\eeq
In the attractive case, $|\kvec - \kvec'|$ approaches $2\Lambda_*$ 
from above as $g\to - \infty$,  and thus corresponds to the 
$+$ sign above.   The $-$ sign then corresponds to the repulsive case
where $2\Lambda_*$ is approached from below.

For two-component fermions,   the phase space factors $\CI$ in
\cite{PyeTon} are doubled,  and since $G\propto 1/\CI$, the
kernels have an extra $1/2$ in the fermionic case:
\beq
\label{fermG}
G_{\rm fermi} =  \inv{2}  G_{\rm bose} 
\eeq

The above unitary limit of the kernels leads to a scale-invariant
integral equation for the pseudo-energy,  which in turn leads
to the scaling forms of the previous section.   This will
be described in detail for the $d=2$ case in subsequent sections.
Note that the kernel has a well-defined expansion in $1/g$;
in the sequel we only consider the above leading terms.    
Solving the integral equation for intermediate values of 
$g$ between $0$ and $\pm \infty$ and inputting the solution
into the expression for the free energy should track the RG flow
between $g=0^-$ and $-\infty$,  or between $g= +\infty $ and
$0^+$,  however we will  not study this in the present work.

\section{Attractive  fermions in 2 dimensions}

\def\zd{y}

The two-dimensional case is considerably simpler to analyze since
the kernel $G$ is a constant.   In any dimension it is convenient
to define:
\beq
\label{2d.1}
\delta (\kvec ) \equiv \vep (\kvec ) - \omega_\kvec + \mu ; ~~~~~~~~~
\zd (\kvec  ) \equiv e^{-\beta \delta (\kvec )} 
\eeq
The expression (\ref{freefoam}) for the free energy can be 
simplified to the following:
\beq
\label{freefoam2}
\CF = -\inv{\beta}  \int (d\kvec) \[ 
- s  \log ( 1- s e^{-\beta \vep  } ) 
-\inv{2}  \frac{ (1-y^{-1} )}{e^{\beta \vep} -s}   \]  
\eeq

Consider first a hypothetical single component fermion. 
In order to have a point-like local interaction one needs
at least two components,  and this will be treated in section VIII.  
As explained there, due to the SU(2) symmetry,  the two-component
fermion reduces to two identical copies of the following 
1-component results.   
Since the kernel is independent if $\kvec$,  $\delta$ is a constant.
The integral equation then becomes the transcendental  algebraic 
equation:
\beq
\label{2d.2}
\zd =  1 \mp  \frac{1}{\zd}  \log ( 1 + z \zd)
\eeq
(We have used $- \Li_1 (-z) = \log (1+z)$.)   
The $-$ ($+$) sign corresponds to the repulsive (attractive) case.  
The free energy takes the form eq. (\ref{freec}) where the
scaling function, expressed in terms of the fugacity  $z$, is  
\beq
\label{2d.3}
c = - \frac{6}{\pi^2} \( 
\Li_2 (-z\zd) + \inv{2} ( 1- y^{-1}) \log (1+ z\zd ) \) 
\eeq
The scaling function $b$ in eq. (\ref{tzero}) is
\beq
\label{2d.3b}
b =  \frac{\pi^2}{3 \log^2 z } c 
\eeq
and the  density is  
\beq
\label{2d.4}
n =  \frac{mT}{2\pi} \log (1 + z\zd)
\eeq
The energy per particle scaling function $\xiprime$  is also
expressed in terms of $c$:
\beq
\label{2d.5}
\xiprime  =  \frac{ 2 \log 2 }{\log ( 1 + z\zd)}c  
\eeq
Finally $\xi$ takes the form:
\beq
\label{2d.5b}
\xi  =  \( \frac{\log z}{\log (1+zy )}\)^2 b
\eeq
We will mainly plot these quantities against $\mu/T$ or its inverse;
these functions are implicitly functions of the more physical
quantity $T/T_F$,  where $T_F = \ep_F/k_B$ is the Fermi temperature.
The relation with $\mu/T$ is
\beq
\label{TTF}
\frac{T}{T_F}  =  \inv{ \log ( 1+ zy)}
\eeq

\begin{figure}[htb] 
\begin{center}
\hspace{-15mm}
\psfrag{Y}{$n/mT$}
\psfrag{X}{$\mu/T$} 
\includegraphics[width=10cm]{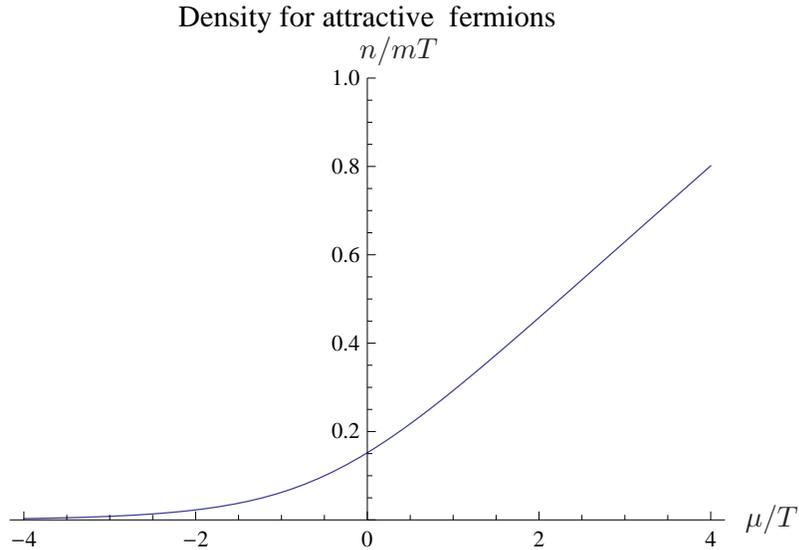} 
\end{center}
\caption{Density of the attractive  fermionic case as a function of
$\mu/T$.}  
\vspace{-2mm}
\label{AtFermDensity} 
\end{figure}

\begin{figure}[htb] 
\begin{center}
\hspace{-15mm}
\psfrag{Y}{$f$}
\psfrag{X}{$\beta \, \kvec^2 / 2m $} 
\psfrag{a}{$\mu / T = 2$}
\psfrag{b}{$\mu / T = 6$}
\includegraphics[width=10cm]{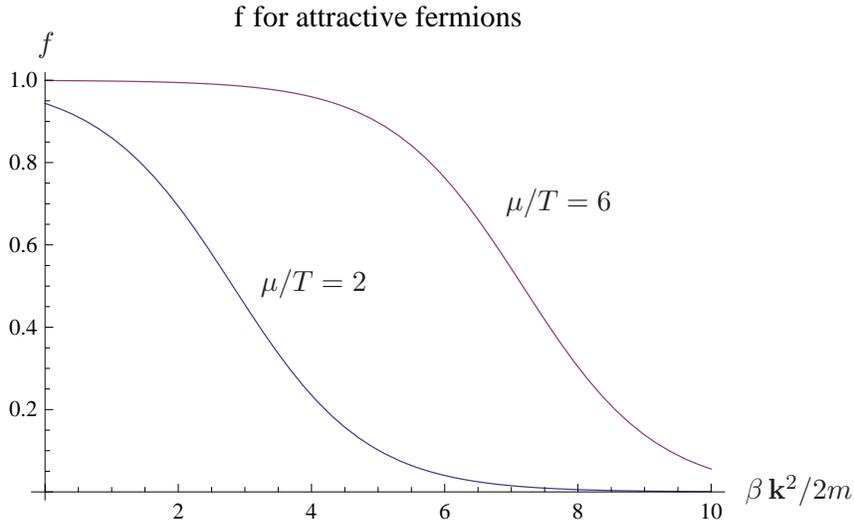} 
\end{center}
\caption{Filling fraction $f$ for  attractive fermions at 
two different values of $\mu / T$. }
\vspace{-2mm}
\label{fAtFermi} 
\end{figure}

\begin{figure}[htb] 
\begin{center}
\hspace{-15mm}
\psfrag{Y}{$c$}
\psfrag{X}{$\mu/T$} 
\includegraphics[width=10cm]{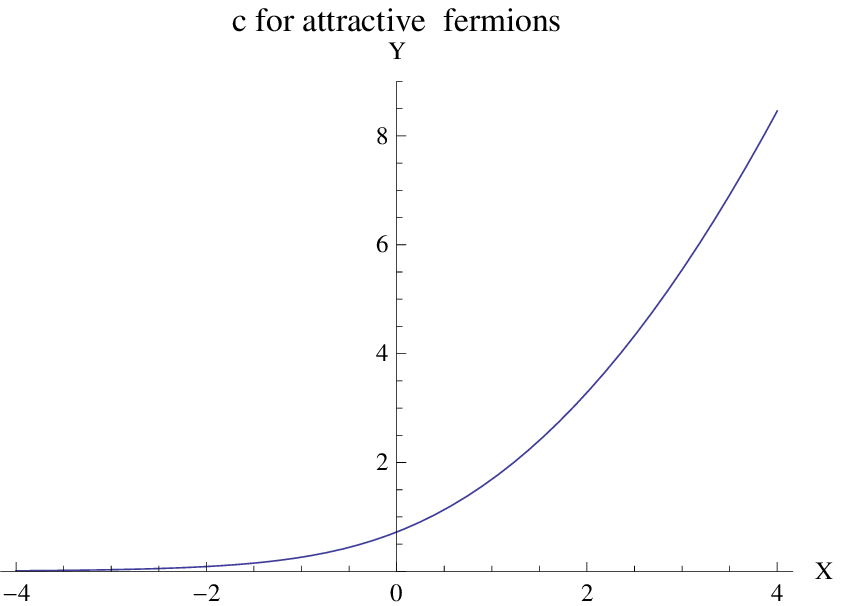} 
\end{center}
\caption{The scaling function $c$  for  attractive  fermions
as a function of
$\mu/T$.}  
\vspace{-2mm}
\label{AtFermc2} 
\end{figure}

There exists a solution to eq. (\ref{2d.2}) for any $z$,  i.e.
for all $-\infty < \mu / T < + \infty$.   The density is shown 
in Figure \ref{AtFermDensity} and takes on all positive values
and approaches $\infty$ as $\mu/T \to \infty$. 
As $T / \mu \to 0^+$,  the density approaches the free
field value $n= m\mu / 2 \pi$.  
However at high temperatures there is a departure from the
free field value:
\beq
\label{denzero}
\lim_{\mu / T \to 0}  n =  0.9546 \frac{mT}{2\pi}
\eeq
The filling fraction $f$ is plotted in Figure \ref{fAtFermi}.     
The scaling function $c$ is shown in Figure \ref{AtFermc2}. 
One finds:
\beq
\label{atferm.1}
\lim_{\mu / T \to 0} c  =  0.624816
\eeq
which is higher than the free field value $c=1/2$.     From these results,  one obtains
the equation of state:
\beq
\label{state.1}
p = 1.077\,  n T, ~~~~~~~~~~~(\mu/T \to 0) 
\eeq

\begin{figure}[htb] 
\begin{center}
\hspace{-15mm}
\psfrag{Y}{$s/m\mu$}
\psfrag{X}{$T/ \mu$} 
\includegraphics[width=10cm]{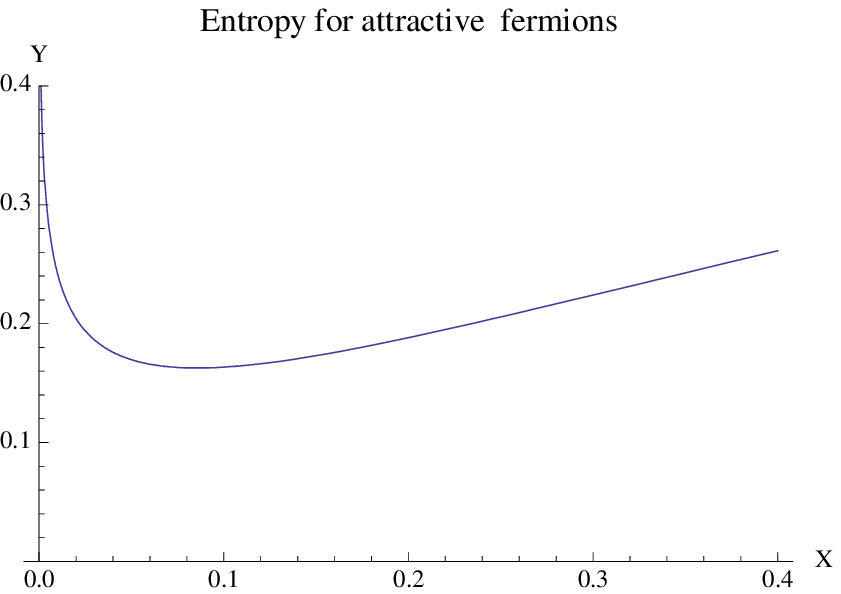} 
\end{center}
\caption{The entropy density $s$ divided by $m\mu$ for  attractive  fermions
as a function of
$T/ \mu$.}  
\vspace{-2mm}
\label{AtFermEntropy} 
\end{figure}

One needs to make sure  that all regions of $\mu/T$   are physical,  and not 
for instance metastable.     In particular,  the entropy must increase with temperature,
otherwise the specific heat is negative.     In Figure  \ref{AtFermEntropy} we plot
the entropy density $s$,   and one sees that  $\d s/ \d T <0$ when $\mu/T > 11.7$.   
One must bear in mind that our formalism,  though non-perturbative in some respects,
is still an approximation,  and such a feature could be an artifact that disappears
once the corrections are incorporated.   Since it is beyond the scope of this work
to systematically explore these corrections,  we will instead try and interpret 
our results as they are,  for this  and subsequent cases.    The above feature of the entropy density 
 could signify a phase transition at  $\mu/T = 11.7$,  where the density
and temperature are related by:
\beq
\label{117}
n =  13.1 \,  \frac{mT}{2\pi}  
~~~~~~\Longleftrightarrow  ~~~~~  T_c  = 0.076\,  T_F
\eeq
where $T_F$ is the Fermi temperature $\ep_F / k_B$,   where $\ep_F$ is defined in
section III,   and $T_c$   
is the critical temperature for this hypothetical phase transition. 
The above value is comparable to the prediction $T_c \approx 0.1 \, T_F$ in \cite{Petrov2}  for 
quasi 2d  (trapped) systems.      In all other regions,  and for all other cases considered below,  
$\d s/ \d T >0$.


\begin{figure}[htb] 
\begin{center}
\hspace{-15mm}
\psfrag{Y}{$b$}
\psfrag{X}{$T/ \mu$} 
\includegraphics[width=10cm]{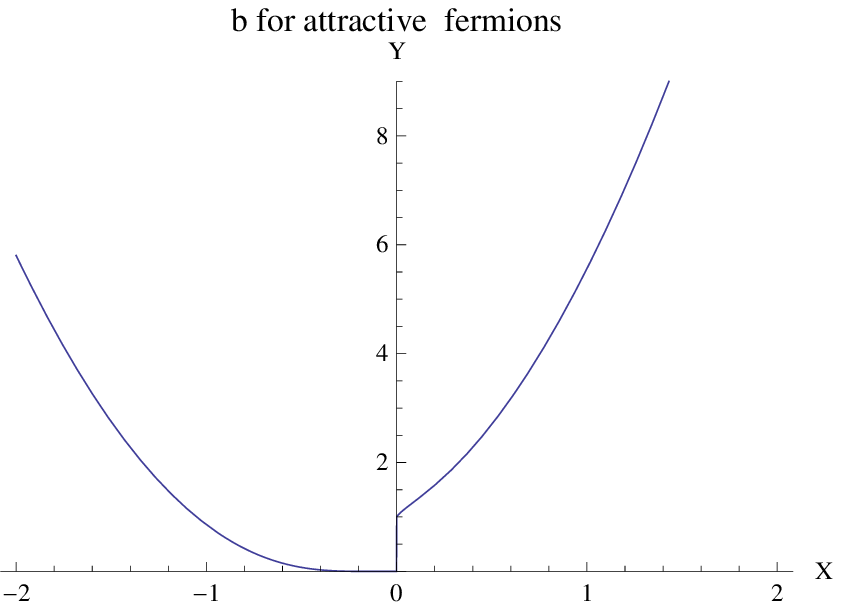} 
\end{center}
\caption{The scaling function $b$  for  attractive  fermions
as a function of
$T/ \mu $.}  
\vspace{-2mm}
\label{AtFermb2} 
\end{figure}

The scaling function $b$ is shown in Figure \ref{AtFermb2}.  
It has a well defined limit as $T / \mu \to 0$, however the
limiting value depends on whether $\mu$ is positive or negative:
\beq
\label{atferm.2}
\lim_{T/\mu \to 0^- } b  = 0, ~~~~~~
\lim_{T/\mu \to 0^+}  b =  1
\eeq
Recall $b = 1$ is the free fermion value.   The discontinuity
at $T/\mu = 0$  is traced to the fact that $c(\infty) \neq c(-\infty)$,
and has no physical significance.    Since the density approaches
$m\mu/2\pi$,    the equation of state is
\beq
\label{state.2}
p  =  2 \pi  n^2  /m   , ~~~~~~~(\mu/T \to \infty)
\eeq

\begin{figure}[htb] 
\begin{center}
\hspace{-15mm}
\psfrag{Y}{$\xiprime$}
\psfrag{X}{$\mu/T$} 
\includegraphics[width=10cm]{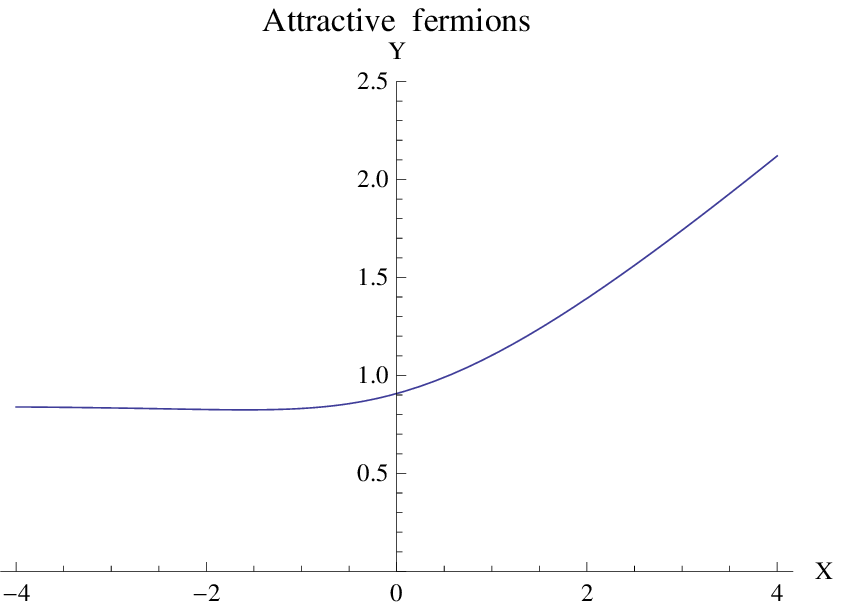} 
\end{center}
\caption{The energy per particle scaling function 
 $\xiprime$  for  attractive  fermions
as a function of
$\mu/T$.}  
\vspace{-2mm}
\label{AtFermXiPrime} 
\end{figure}

The energy per particle scaling function $\xiprime$ is shown
in Figure \ref{AtFermXiPrime}.   It has the limiting values:
\beq
\label{atfermXiprime}
\lim_{\mu/T \to 0} \xiprime  = 0.907412, 
~~~~~
\lim_{\mu/T \to - \infty } \xiprime  =0.842766, 
~~~~~~~
\lim_{\mu/T \to \infty}  \xiprime  = \infty
\eeq
The above value for $\xiprime$ as $\mu/T \to -\infty$ 
was determined numerically,  however it turns out 
to equal $12 \log 2/\pi^2$.   Equation (\ref{enpart.5}) 
then shows that the energy per particle  $\ep_1 = T$,
which simply means that in the limit $\mu/T \to -\infty$ 
the gas is effectively classical.    This feature will
be encountered in other cases below.

\begin{figure}[htb] 
\begin{center}
\hspace{-15mm}
\psfrag{Y}{$\xi$}
\psfrag{X}{$T/ \mu$} 
\includegraphics[width=10cm]{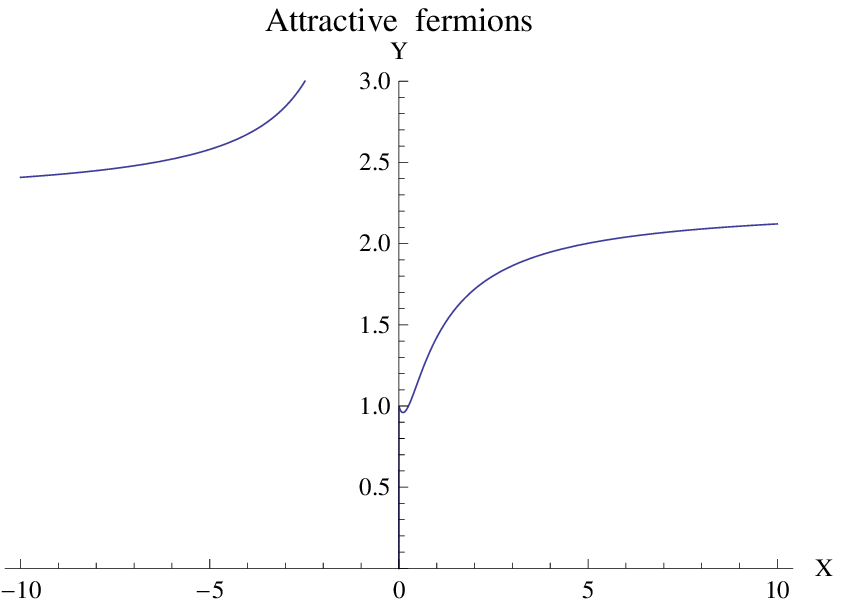} 
\end{center}
\caption{The energy per particle 
scaling function $\xi $  for  attractive  fermions
as a function of
$T/ \mu $.}  
\vspace{-2mm}
\label{AtFermXiTmu} 
\end{figure}

The single particle energy scaling function $\xi$ has
the  behaviour shown in Figure \ref{AtFermXiTmu}. 
It has the limiting behavior
\beq
\label{atfermxi}
\lim_{T/ \mu \to 0^+} \xi  =  1, ~~~~~~~
\lim_{T/\mu \to \pm \infty} \xi = 2.25593
\eeq
The value $\xi (0)  = 1$ is consistent with the arguments 
in \cite{Nussinov}.  
It diverges as $T/\mu \to 0^-$ due to the vanishing density
at $\mu / T \to -\infty$.

\def\lfree{\ell_{\rm free} } 
\def\vbar{\bar{v}}

\section{Ratio of shear viscosity to entropy density}

The ratio of the shear viscosity $\eta$ to the entropy density $s$
has units of  $\hbar / k_B$ in any dimension.   A lower bound
was conjectured for  3d  relativistic systems\cite{Kovtun}:
\beq
\label{bound}
\frac{\eta}{s} \geq  \frac{\hbar}{4\pi k_B}
\eeq
based on the AdS/CFT correspondence.  The bound is saturated in
certain strongly coupled supersymmetric gauge theories.  
It is therefore interesting to study this ratio 
for non-relativistic,  strongly interacting systems, in 3 and lower dimensions,  
since it is unknown whether there really is a lower bound. 
This ratio was studied for $3d$ 
unitary Fermi gases in \cite{Massignan,Gelman,Schafer,Rupak}.
The attractive fermion case is the most interesting
and well-behaved case in our formalism, as will be evident
in the subsequent sections,  so we study the issue in  this case first.  

Consider first a gas with a single species of particle of mass $m$.  
For a non-relativistic system in 2 spatial dimensions:
\beq
\label{eta.1} 
\eta =  \inv{2}   n  \bar{v} m \lfree
\eeq
where $\vbar$ is the average speed and $\lfree$ the
mean free path.    The mean free path is 
$\lfree = 1/(\sqrt{2} n \sigma)$ where $\sigma$ is the total
cross-section.  The $\sqrt{2}$ comes from the ratio of the mean
speed to the mean relative speed\cite{Reif}.  
In the unitary limit,  formally $S = 1 + i \CI \CM = -1$,
which implies the scattering amplitude $\CM = 2 i / \CI$.  
(See section IV.)    
The cross-section in eq. (\ref{cross}) gives 
$\sigma = m^2 / k \CI^2$ in 2 dimensions,  where
$\CI = m/4$\cite{PyeTon}.   Thus,
in the unitary limit 
\beq
\label{eta.2}
\sigma =  \frac{16}{|\kvec|}
\eeq
where $\kvec$ is a single particle momentum.   This gives
\beq
\label{eta.3}
\eta =  \frac{m}{16\sqrt{2}} \(  \inv{2} m  \bar{v}^2  \)
\eeq
Since the pressure is determined by the average kinetic energy,
and the ideal gas relation eq. (\ref{ener}) still holds 
for a unitary gas, the average kinetic energy per particle
is just $\ep_1 = -\CF/n$, eq. (\ref{enpart.1}).    Finally we obtain:
\beq
\label{eta.4} 
\frac{\eta}{s} =  \frac{m}{16 \sqrt{2}} \frac{\CF}{n}  
\(  \frac{  \d \CF }{\d T} \)^{-1} 
\eeq
where all the quantities in the above formula are the single
component values of the last section.  In terms of the scaling
functions:
\beq
\label{eta.5}
\frac{\eta}{s}  =  \frac{ 3  }{ 4\sqrt{2} \pi} \frac{c}{ c' }  \frac{~}{( 2 c - \frac{\mu}{T} c' )}
=  \frac{ \pi}{4 \sqrt{2} } \frac{b}{b' } \frac{~}{( 2b - \frac{T}{\mu} b' )}
\eeq

 For two-component fermions the 
available phase space 
$\CI$ is doubled.   The cross-section is halved since
spin up particles only scatter with spin down.   Finally the entropy is doubled.
This gives
\beq
\label{etasbosfer}
\( \frac{\eta}{s}  \) _{\rm fermi}  =  4 \(\frac{\eta}{s} \)_{\rm bose}
\eeq

\begin{figure}[htb] 
\begin{center}
\hspace{-15mm}
\psfrag{Y}{$\eta/s$}
\psfrag{X}{$\mu/T$} 
\includegraphics[width=10cm]{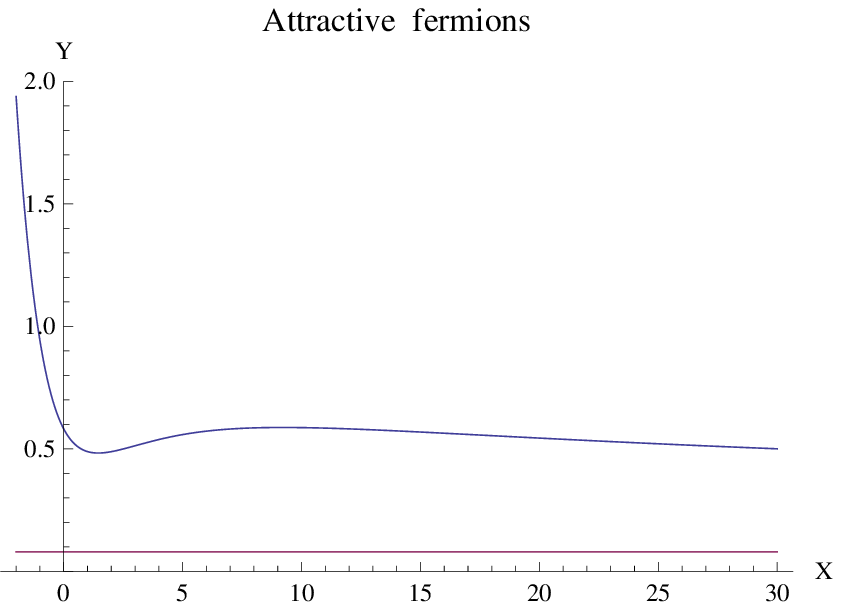} 
\end{center}
\caption{The ratio $\eta/s$ as a function of $\mu/T$.
The horizontal line is $1/4\pi$.}  
\vspace{-2mm}
\label{etassmall} 
\end{figure}

\begin{figure}[htb] 
\begin{center}
\hspace{-15mm}
\psfrag{Y}{$\eta/s$}
\psfrag{X}{$T/T_F$} 
\includegraphics[width=10cm]{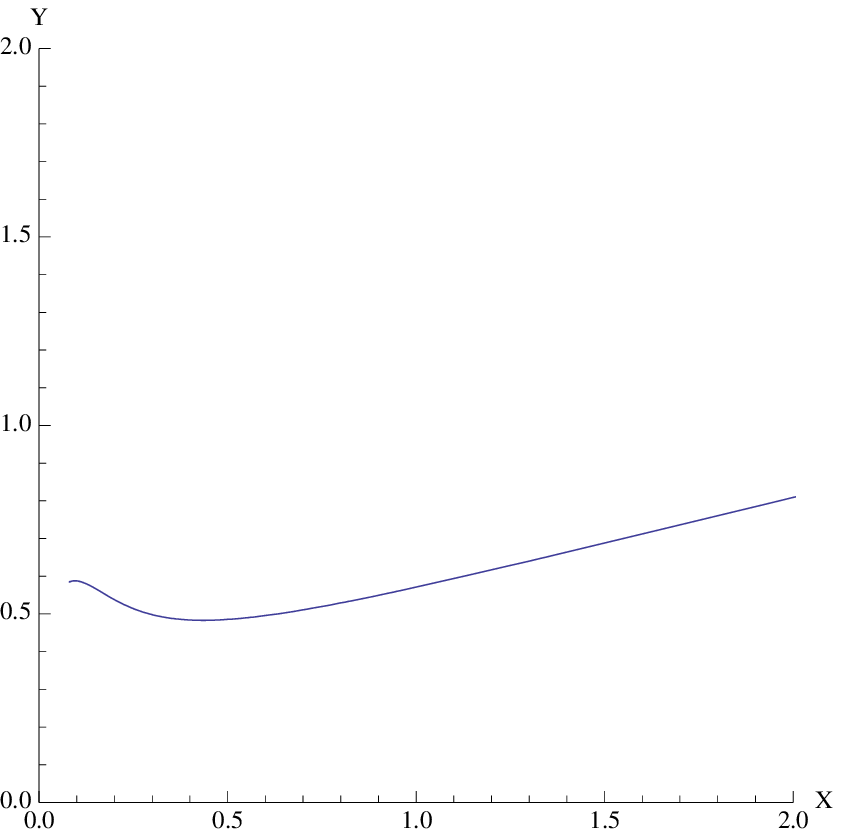} 
\end{center}
\caption{The ratio $\eta/s$ as a function of $T/T_F$ for attractive fermions.}
\vspace{-2mm}
\label{EtasTTF} 
\end{figure}

The above expression is easily evaluated numerically using
the expressions of the last section.   The result is displayed
for small values of $\mu/T$ in Figure \ref{etassmall}.   
In this regime,  $\eta/s$ is well above the conjectured bound,
and comparable to the $3d$ values extracted 
from the experimental data\cite{Schafer}.   
We find
\beq
\label{etaslim.1}
\lim_{\mu/T \to 0}  \frac{\eta}{s} =  7.311 \frac{\hbar}{4 \pi k_B} 
\eeq
This is well-below the values for common substances like water,
however it is comparable to values for liquid helium, which is
about 9 times the bound\cite{Kovtun}.    In the region shown:
\beq
\label{etasmin}
\frac{\eta}{s}  \geq 6.07   \frac{\hbar}{4 \pi k_B}
\eeq
In Figure \ref{EtasTTF}  we plot $\eta/s$ as a function of $T/T_F$,
and one observes a behavior quite similar both qualitatively and quantitatively to the 3d data 
summarized  in\cite{Schafer},    where the minimal value is about 6 times the bound.

\begin{figure}[htb] 
\begin{center}
\hspace{-15mm}
\psfrag{A}{$\inv{4 \pi}$}
\psfrag{Y}{$\eta/s$}
\psfrag{X}{$\mu/T$} 
\includegraphics[width=10cm]{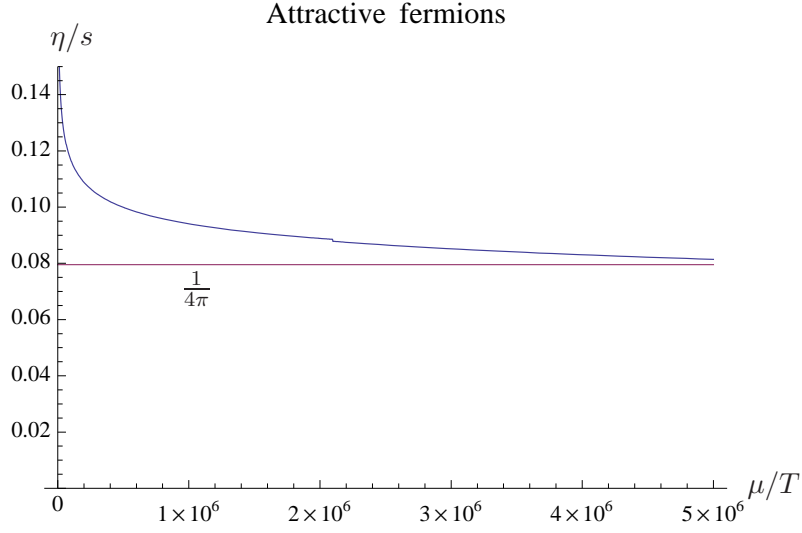} 
\end{center}
\caption{The ratio $\eta/s$ as a function of $\mu/T$ 
as $\mu/T$ gets very large..
The horizontal line is $1/4\pi$.}  
\vspace{-2mm}
\label{EtasMiddle} 
\end{figure}

\begin{figure}[htb] 
\begin{center}
\hspace{-15mm}
\psfrag{A}{$\inv{4 \pi}$}
\psfrag{Y}{$\eta/s$}
\psfrag{X}{$\mu/T$} 
\includegraphics[width=10cm]{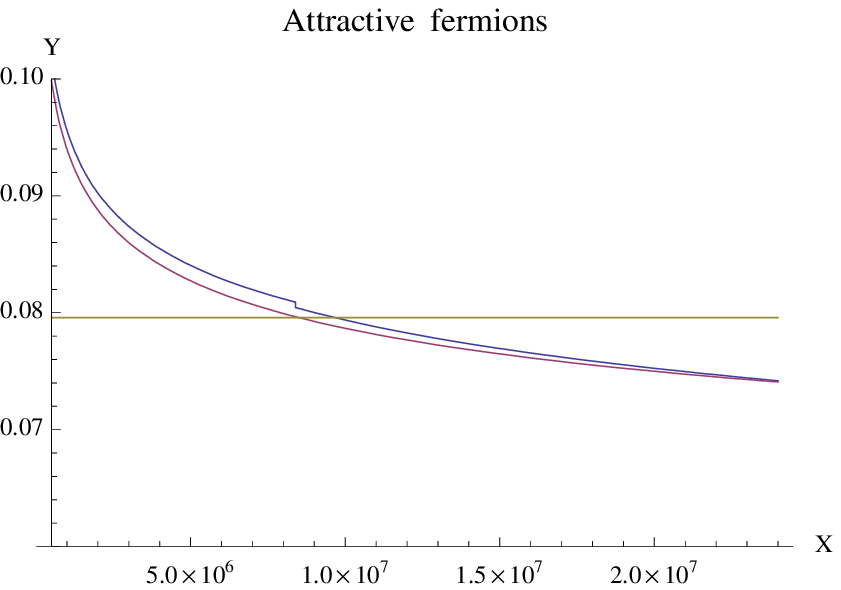} 
\end{center}
\caption{The ratio $\eta/s$ as a function of $\mu/T$ 
in the limit of very low temperatures (top curve). 
The horizontal line is $1/4\pi$. The bottom curve is the 
approximation eq. (\ref{etasap}).}  
\vspace{-2mm}
\label{EtasLarge} 
\end{figure}

Recall that in the last section,   it was argued that the region $\mu/T >  11.7$ is
unphysical, or perhaps metastable,  since the entropy increases with decreasing temperature there. 
Thus for the regions that are surely physical,  the bound (\ref{etasmin}) holds.  
It is nevertheless interesting to study the viscosity in this  unphysical region.    
In the zero temperature limit, i.e. $T/\mu \to 0^+$, 
$\eta/s$  slowly decreases and seems to approach the bound.
See Figure \ref{EtasMiddle}.   However it eventually dips below it.     
See Figure \ref{EtasLarge}. 
This behavior can be understood analytically as follows.
  For $x=\mu/T$ very large,  the solution
to the equation (\ref{2d.2}) is approximately:
\beq
\label{yap}
y(x) \approx  \sqrt{x}   + 1/2  + (\log x - 1 )/4\sqrt{x}
\eeq
This leads to the asymptotic expansions:
\barray
\label{cnap}
c &\approx&  3 (x^2 + x\log x -x )/\pi^2  
\\   \nonumber
n &\approx&  \frac{mT}{2\pi}  (x + \log(x)/2 )
\earray
and 
\beq
\label{etasap}
\frac{\eta}{s}  \approx  \frac{\pi}{2\sqrt{2}}\(   1/( \log (\mu/T)  -2 )  + T/2\mu   \)  
\eeq
In terms of the density:
\beq
\label{etaslim}
\lim_{\mu/T \to \infty}  \frac{\eta}{s} 
= \frac{\pi}{2\sqrt{2}  \log (2 \pi n/ mT)}
\eeq

It is important to note that  although the energy per particle scaling function $\xi$  approaches
the free field value at low temperatures,   the above behavior is very different from  the free field case.
In the latter,  the scaling function $c= - 6 \Li_2 (-z)/\pi^2$,   which gives the diverging
behavior at very low temperature:
\beq
\label{etasfree}
\frac{\eta}{s}  \approx  \frac{3}{4 \sqrt{2} \pi}  \frac{\mu}{T} , ~~~~~~~~~~({\rm free ~ fermions})
\eeq

Finally, as $\mu/T \to - \infty$,  $y\approx 1+z$,
and $c\approx 6 z/\pi^2$ and $n\approx mT z/2\pi$.
This leads to exponential growth:
\beq
\label{etaslimmin}
\lim_{\mu/T \to - \infty}  \frac{\eta}{s}  =   -  \frac{\pi T}{2\sqrt{2} \mu} 
  e^{-\mu/T}  =  - \frac{mT}{4\sqrt{2}} \inv{ n \log (2\pi n/mT)}
\eeq

\section{Attractive Bosons in 2 dimensions}

Using the same definitions as for fermions in eq. (\ref{2d.1}),
the integral equation for bosons is 
\beq
\label{bos.1}
y =  1 \pm \frac{2}{y} \log ( 1- z y) 
\eeq
where the $+$ ($-$) sign refers to repulsive (attractive) interactions.  
The scaling function $c$ is now
\beq
\label{bos.c2}
c =  \frac{6}{\pi^2} \( \Li_2 (zy)  + \inv{2} (1-y^{-1}) \log (1-zy) \)
\eeq
The scaling function $b$ has the same expression as in
eq. (\ref{2d.3b}), with the above $c$.     
The density now  is
\beq
\label{bos.density}
n = - \frac{mT}{2\pi} \log (1 - z y) 
\eeq
The energy per particle scaling functions  now take  the form:
\beq
\label{bos.xiprime}
\xiprime  =  -  \frac{2 \log (2) c  }{\log(1-zy) }
\eeq
and 
\beq
\label{xibosdef}
\xi =  \(  \frac{\log z}{\log (1-zy)} \)^2  b
\eeq

\def\zmax{z_c}

\begin{figure}[htb] 
\begin{center}
\hspace{-15mm}
\psfrag{Y}{$n/mT$}
\psfrag{X}{$\mu/T$} 
\includegraphics[width=10cm]{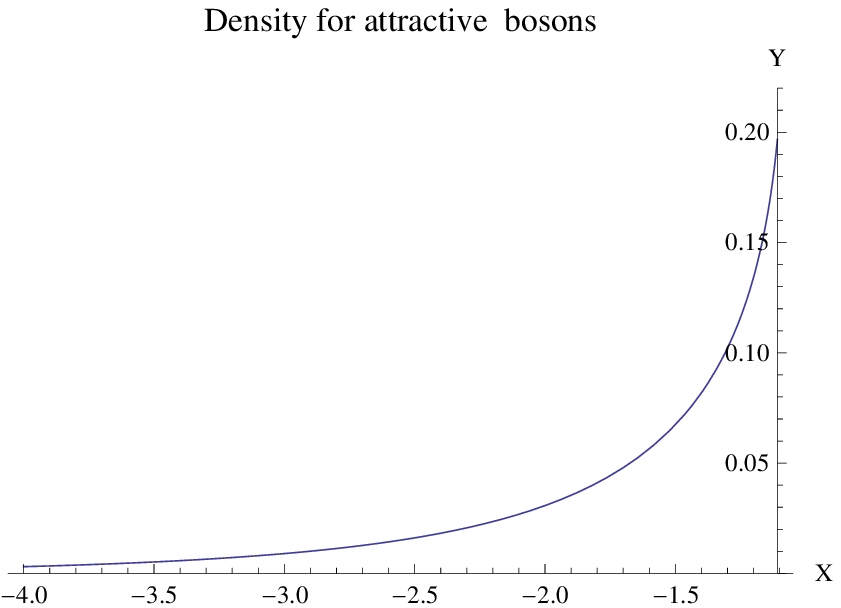} 
\end{center}
\caption{Density of the attractive   bosons as a function of
$\mu/T$. The origin of the axes is at $(\log \zmax , 0)$.}  
\vspace{-2mm}
\label{AtBosDensity} 
\end{figure}

For this bosonic case, there is only a solution to eq. (\ref{bos.1}) for 
$z \leq \zmax \approx  .34$,  or $\mu/T \leq -1.08$.   
The density is shown in Figure \ref{AtBosDensity}, and note that it
has a maximum.   
The filling fractions are shown for several $\mu /T$ up to
$\log \zmax$.     From these plots, it appears that $f$ is
diverging at $\kvec =0$ as $z$ approaches $\zmax$,  suggestive
of condensation to a superfluid.   
Let us refer  to the maximum density then as the critical density
\beq
\label{atbos.den}
n_c  \approx   1.24 \,  \frac{m \, k_B T}{2 \pi \hbar^2}
~~~~ \Longleftrightarrow  ~~~~ T_c \approx 0.81 \, T_F
\eeq


\begin{figure}[htb] 
\begin{center}
\hspace{-15mm}
\psfrag{Y}{$f$}
\psfrag{X}{$\beta \kvec^2 /2m $} 
\psfrag{a}{$\mu /T = \log \zmax $} 
\includegraphics[width=10cm]{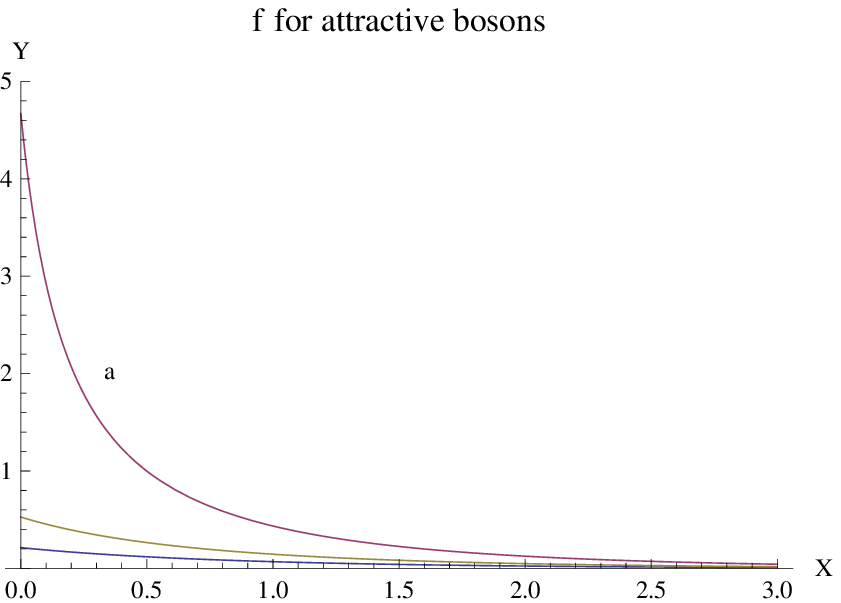} 
\end{center}
\caption{The filling fraction $f$ for attractive bosons 
for $\mu/T = -2,-1.5, -1.08=\log \zmax$. }  
\vspace{-2mm}
\label{fAtBos} 
\end{figure}

\def\zmax{z_c}

\begin{figure}[htb] 
\begin{center}
\hspace{-15mm}
\psfrag{Y}{$c$}
\psfrag{X}{$\mu/T$} 
\includegraphics[width=10cm]{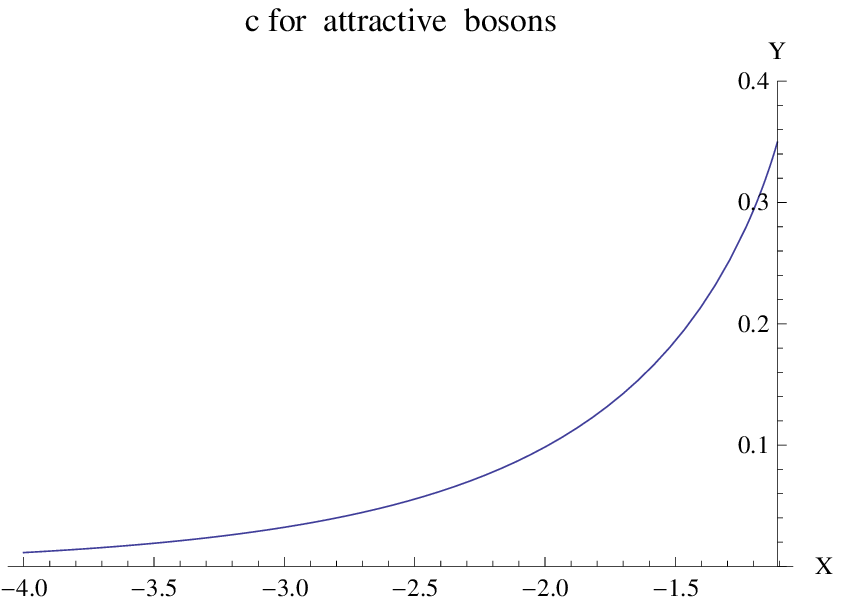} 
\end{center}
\caption{The scaling function $c$  for  attractive  bosons 
as a function of
$\mu/T$.  The origin of the axes is at $(\log .33, 0)$.}  
\vspace{-2mm}
\label{AtBosc2} 
\end{figure}

The scaling function $c$ is shown in Figure \ref{AtBosc2}.  
The limiting value is 
\beq
\label{atbos.c2}
 c (\zmax )  \approx 0.35
\eeq
which is considerably less than for a free boson with $c = 1$.  
The function $b$ decreases to zero at $T/\mu \to 0$, and 
the limiting value is
\beq
\label{atbos.b2}
 b (\zmax )  \approx 0.94
\eeq

\begin{figure}[htb] 
\begin{center}
\hspace{-15mm}
\psfrag{Y}{$\xiprime$}
\psfrag{X}{$\mu/T$} 
\includegraphics[width=10cm]{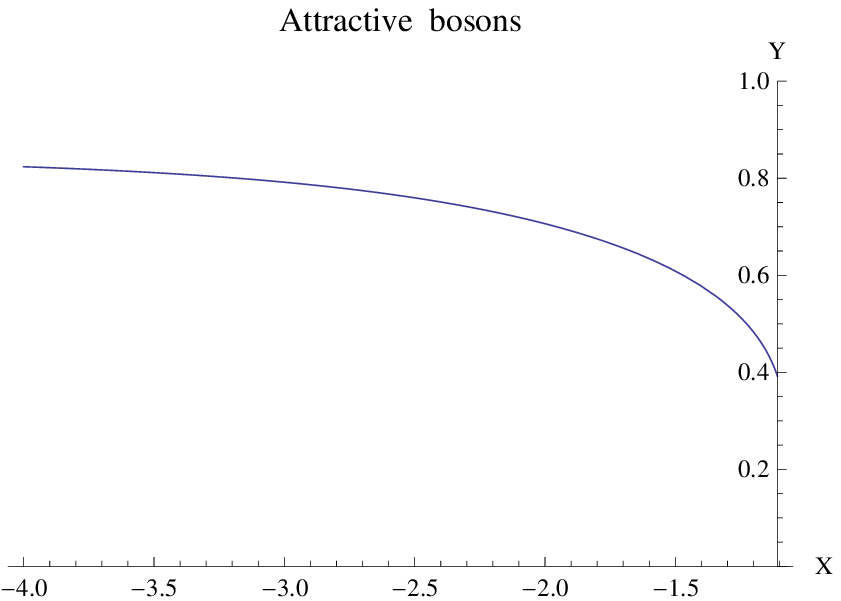} 
\end{center}
\caption{The energy per particle scaling function 
 $\xiprime $  for  attractive    bosons 
as a function of
$\mu/T$.}  
\vspace{-2mm}
\label{AtBosXiPrime} 
\end{figure}

\begin{figure}[htb] 
\begin{center}
\hspace{-15mm}
\psfrag{Y}{$\xi$}
\psfrag{X}{$T/ \mu$} 
\includegraphics[width=10cm]{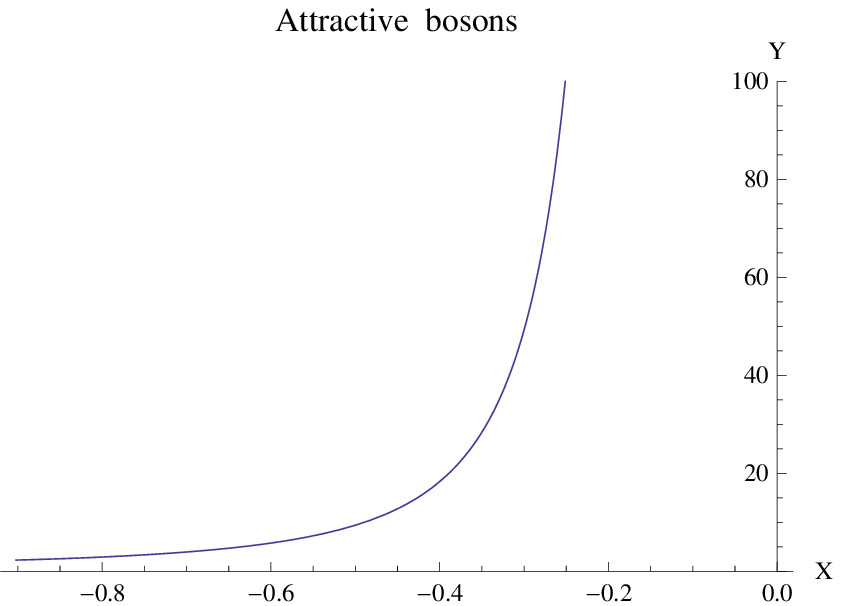} 
\end{center}
\caption{The energy per particle 
 scaling function $\xi $  for  attractive   bosons 
as a function of
$T/ \mu $.}  
\vspace{-2mm}
\label{AtBosXiTmu} 
\end{figure}

The energy per particle scaling functions  $\xi$ and $\xiprime$ are shown
in Figures \ref{AtBosXiPrime}, \ref{AtBosXiTmu}. 
The limiting values are
\beq
\label{atbos.xip}
\lim_{\mu/T \to - \infty}  \xiprime = 0.842766, ~~~~~~
\xiprime (\zmax ) \approx 0.39
\eeq
and 
\beq
\label{atbos.xi}
\lim_{T/\mu \to 0^-} \xi = \infty,  ~~~~~~
\xi (\zmax) \approx 2.3
\eeq
As for the fermionic case,  the value $\xiprime =0.842766 = 12 \log 2/\pi^2$
implies the energy per particle $\ep_1 = T$, which means the gas is
in the classical limit.

The ratio  $\eta/s$ given by the expressions eq. (\ref{eta.5}) 
with the appropriate bosonic $c$ and $b$.   
We find that $\eta/s$ has its minimum value at $z_c$, where it
is actually below the bound
\beq
\label{etasboszc}
\frac{\eta}{s} \geq   0.4\,  \frac{\hbar}{4\pi k_B}
\eeq


\section{Repulsive fermions and bosons in 2 dimensions}

\subsection{Fermions} 

\begin{figure}[htb] 
\begin{center}
\hspace{-15mm}
\psfrag{Y}{$n/mT$}
\psfrag{X}{$\mu/T$} 
\psfrag{a}{$n_c$}
\includegraphics[width=10cm]{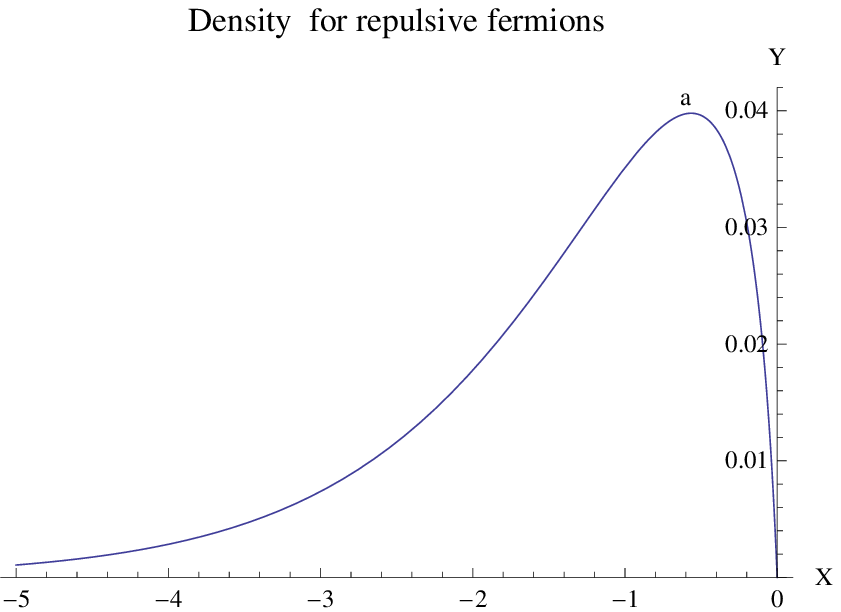} 
\end{center}
\caption{Density of the repulsive  fermionic case as a function of
$\mu/T$.}  
\vspace{-2mm}
\label{RepFermDensity} 
\end{figure}

\begin{figure}[htb] 
\begin{center}
\hspace{-15mm}
\psfrag{Y}{$f$}
\psfrag{X}{$\beta\, \kvec^2 /2m $} 
\psfrag{b}{$n_c$}
\includegraphics[width=10cm]{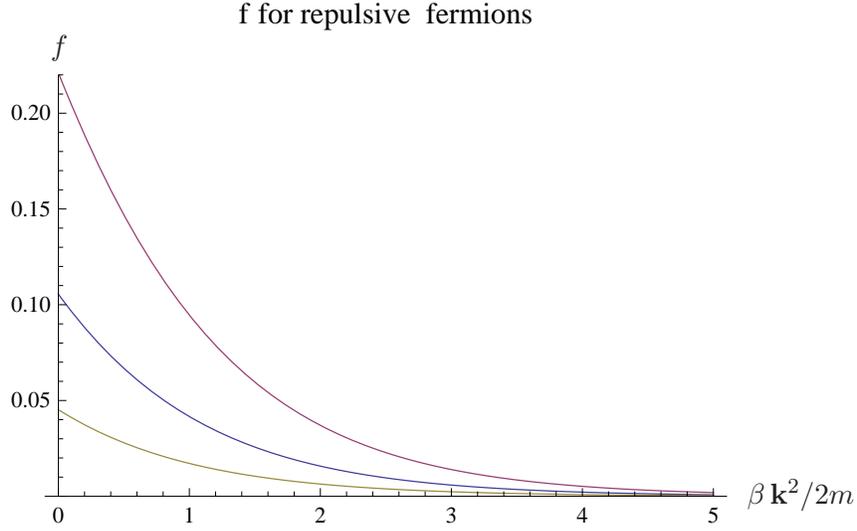} 
\end{center}
\caption{Filling fraction $f$ for  repulsive  fermions 
for $\mu/T = -3, -2, -0.55$. The top curve corresponds to 
the critical density in eq. (\ref{repferm.1}). }  
\vspace{-2mm}
\label{fRepFerm} 
\end{figure}

\begin{figure}[htb] 
\begin{center}
\hspace{-15mm}
\psfrag{Y}{$c$}
\psfrag{X}{$\mu/T$} 
\includegraphics[width=10cm]{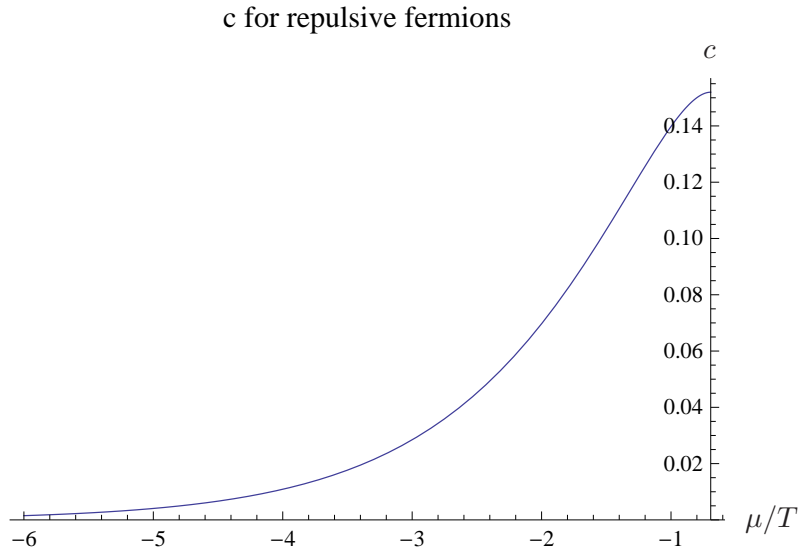} 
\end{center}
\caption{The scaling function $c$  for  repulsive   fermions
as a function of
$\mu/T$.  }  
\vspace{-2mm}
\label{RepFermc2} 
\end{figure}

There are solutions  to the eq. (\ref{2d.2}) for all
$\mu/T < 0$, and the 
   density is positive in this range.  
A plot of the density is shown in Figure \ref{RepFermDensity}. 
The density maximizes at $\mu/T \approx -0.56$, where
$n/mT \approx 0.04$.  
This seems physically reasonable given the strong repulsion
in the unitary limit.  
In contrast,  for small coupling $g$,  the kernel 
$G \approx -g$,  and there are solutions for positive
chemical potential.   In the bosonic case,  our
formalism leads to a critical density of $n_c = \frac{mT}{2\pi} \log(2\pi/mg)$
for the Kosterlitz-Thouless transition\cite{PyeTon}.   
  The filling fractions are shown in
Figure \ref{fRepFerm}.  Note that they are considerably smaller
than in the attractive case,  as expected. 
The scaling function $c$ is shown in Figure \ref{RepFermc2}. 
It has the limiting value
\beq
\label{crepferm}
\lim_{\mu / T \to 0} c = 0.303964,
\eeq
which is considerably less than the free field value
$c = 1/2$.  
The scaling function $b$ 
 has the limiting values
$\lim_{T/\mu \to 0^-} b = 0$, and 
$\lim_{T/ \mu \to -\infty } b =  \infty$.

Since the pressure is proportional to $c$, 
the same density occurs at two different pressures.
One  sees that for $\mu /T < -0.56$,  the density increases
with pressure as it should.   For $\mu /T > -0.56$ the density
decreases with increasing pressure,  violating $dp/dV < 0$,
and should thus be viewed as 
 an unphysical region.        We interpret this as a phase transition
occuring at  $z_c = e^{-0.56} = 0.57$, where the critical density
\beq
\label{repferm.1}
n_c  =  0.25 \,  \frac{m k_B T}{2\pi \hbar^2}
~~~~\Longleftrightarrow  ~~~~ T_c \approx 4.0 \, T_F
\eeq
At this critical point $c(-0.56) = 0.24$. 
This perhaps corresponds to a transition to a crystaline phase.
We cannot prove this,  since we have not calculated the shear
modulus for instance,  and it could simply be an artifact of
our approximation,   but let us take it as a hypothesis.
Whereas a Wigner crystal phase occurs at low density in
three dimensions,  it occurs at high density in two dimensions\cite{crystal1}.
The possibility of crystal phases for repulsive bosons was
studied in \cite{Kolomeisky}.    
For Coulomb repulsion with strength $e^2$, the thermodynamic state of
a classical Coulomb system is determined by 
$\Gamma = \sqrt{\pi n} \, e^2 /k_B T$,  where $n$ is the density. 
Experimentally it was found that $\Gamma \approx 137$\cite{Grimes}, 
which gives a critical density
\beq
\label{Gammacrystal}
n_c  \approx  2.15 \times 10^9  \,  T^2  \, \inv{{\rm cm}^2 K^2}   
\eeq
On the other hand,  since our model has point-like interactions,
and is  in the unitary limit,  the conditions for a crystal phase
are expected  to be different.  In particular 
our formula eq. (\ref{repferm.1})  has no $e^2$ dependence,
and this leads to a linear in $T$ dependence.    For $m$ equal to the
electron mass
\beq
\label{crystalnc}
n_c \approx 4.5 \times 10^9 \,  T  \, \inv{{\rm cm}^2 K}
\eeq
It is interesting to note that for $T$ of order 1, which is where
the data in \cite{Grimes} was taken,    the two
densities (\ref{Gammacrystal}) and (\ref{crystalnc})  
are comparable.

\begin{figure}[htb] 
\begin{center}
\hspace{-15mm}
\psfrag{Y}{$\xiprime$}
\psfrag{X}{$\mu/T$} 
\includegraphics[width=10cm]{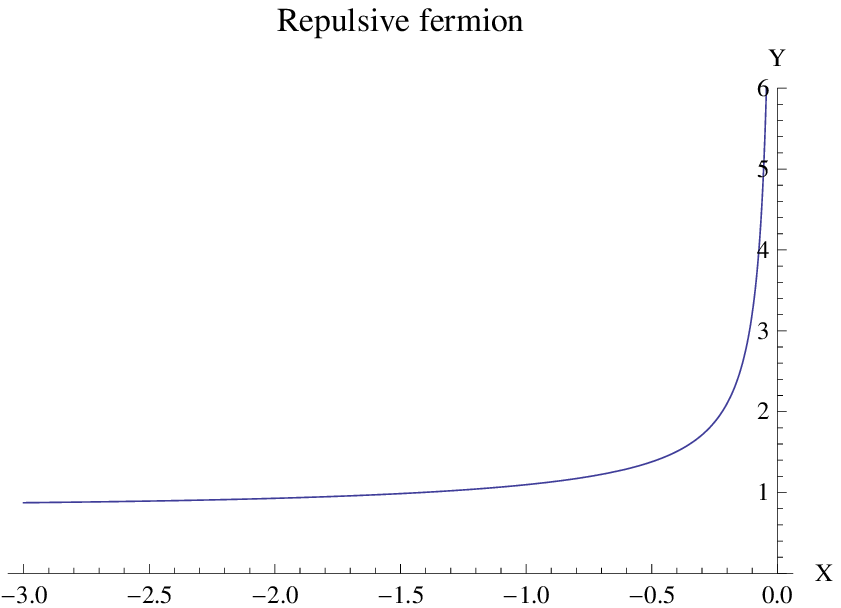} 
\end{center}
\caption{The energy per particle scaling function 
 $\xiprime $  for  repulsive   fermions
as a function of
$\mu/T$.}  
\vspace{-2mm}
\label{RepFermXiPrime} 
\end{figure}

The energy per particle function $\xiprime$ is shown in Figure
\ref{RepFermXiPrime}.   The limiting values are:
\beq
\label{repferm.xiprime}
\lim_{\mu / T \to - \infty} \xiprime = 0.842766, 
~~~~~
\xiprime (z_c) =  1.32
\eeq
As in previous cases,  the above value of $\xiprime$ implies
$\ep_1 =T$ as $\mu/T \to -\infty$ and the gas is thus  in the
 classical limit. 
Beyond the critical point where the density increases, 
$\lim_{\mu / T \to 0 } \xiprime =  \infty$.  
Note that $\xiprime$ starts to diverge around the proposed phase
transition at $\mu/ T =  -0.56$.  
Finally the other energy per particle function $\xi$  
is less interesting, as it diverges at both endpoints 
of the density range where the density goes to zero. 
At the critical point it is still quite large:
$\xi (z_c ) = 12.6$.

The ratio $\eta/s$ reaches its minimum near $z_c$, 
and is quite large compared to the attractive case:
\beq
\label{etasrepf}
\frac{\eta}{s} \geq 22.6\,  \frac{\hbar}{4\pi k_B}
\eeq

\subsection{Repulsive Bosons}

\begin{figure}[htb] 
\begin{center}
\hspace{-15mm}
\psfrag{Y}{$n/mT$}
\psfrag{X}{$\mu/T$} 
\includegraphics[width=10cm]{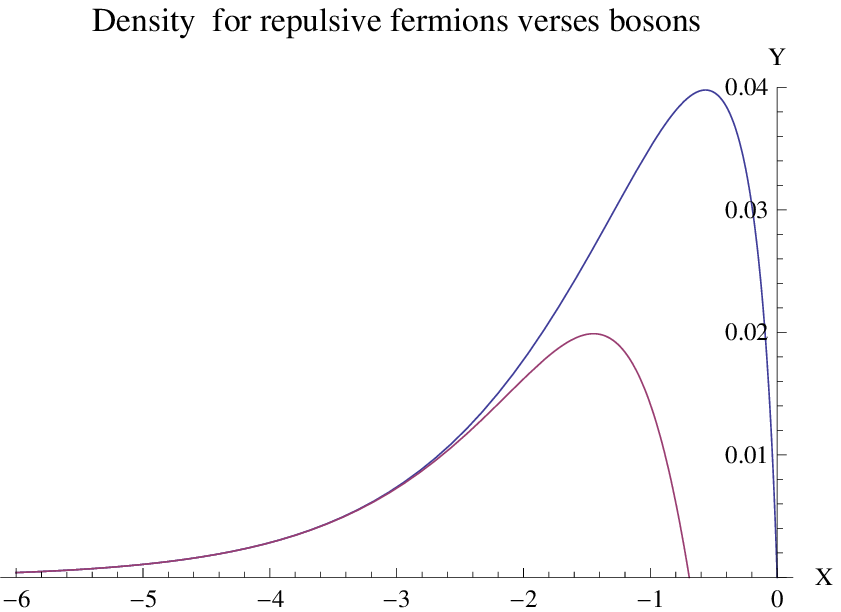} 
\end{center}
\caption{Density of  repulsive  bosons (bottom curve)  verses fermions  as a function of
$\mu/T$.}  
\vspace{-2mm}
\label{BothDensity} 
\end{figure}

The  repulsive boson case is similar  to the
repulsive fermion,  except that now  there are only positive density
solutions to
the eq. (\ref{bos.1}) for $\mu / T < - \log 2$.   
Figure  \ref{BothDensity}
compares   the density  for bosons verses fermions.  
The maximum density for bosons,  which could again possibly signify
a critical point at $z_c \approx e^{-1.45} = 0.235$,  
is half that of the fermion case:
\beq
\label{ncbos}
n_c = 0.125  \, \frac{m k_B T}{2 \pi \hbar^2} ~~~~~
\Longleftrightarrow  ~~~~~T_c \approx 8.0 \, \ep_F
\eeq
At this point $c(-1.3) \approx 0.11$.

\begin{figure}[htb] 
\begin{center}
\hspace{-15mm}
\psfrag{Y}{$\eta/s $}
\psfrag{X}{$\mu/T$} 
\includegraphics[width=10cm]{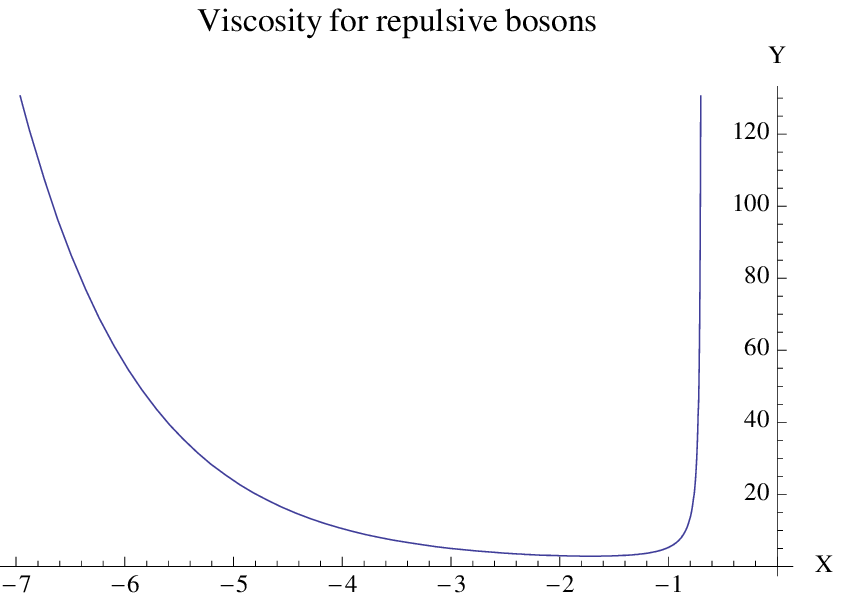} 
\end{center}
\caption{The ratio $\eta/s$  as a function of
$\mu/T$  for repulsive bosons.}  
\vspace{-2mm}
\label{EtasRepBos} 
\end{figure}

The ratio $\eta/s$  is shown in Figure \ref{EtasRepBos}.   It  has a minimum at $\mu/T = -1.72$  
near the critical point and in the physical region:
\beq
\label{etasbosrep}
\frac{\eta}{s}  \geq  8.85 \,   \frac{\hbar}{4\pi k_B}
\eeq
Recall,  the region $\mu/T > -1.45$ is unphysical.

\section{Multiple species of particles}

Let us suppose the gas consists of multiple species
of particles labeled by the index ``a'',  of mass
$m_a$,  chemical potential  $\mu_a$,  and statistical
parameter $s_a = \pm 1$.      Introduce a
pseudo-energy $\vep_a (\kvec )$ for each type of particle
and the filling fractions:
\beq
\label{multi.f}
f_a (\kvec )  =  \inv{ e^{\beta \vep_a (\kvec)} - s_a }
\eeq
As shown in Appendix B,  in our approximation, 
 the pseudo-energies satisfy
the coupled integral equations:
\beq
\label{multi.integral}
\vep_a (\kvec ) = \omega_a (\kvec ) - \mu_a - \inv{\beta} 
\log \(  1 + \beta \sum_b G_{ab} * 
 \frac{y_b^{-1}}{e^{\beta \vep_b}  - s_b} \)
\eeq
where $\omega_a (\kvec ) = \kvec^2/2m_a $, 
\beq
\label{multi.zd}
y_a \equiv e^{-\beta \delta_a},   ~~~~~~~  \delta_a  \equiv 
\vep_a  - \omega_a  + \mu_a 
\eeq
and $G_{ab}$ is the 2-body scattering kernel.   
By Galilean invariance,  the kernel is a function of
$|v_a - v_b|$  where $v = k/m $ is a velocity. 
For any function $g$ we defined:
\beq
\label{multi.Gstar}
\( G_{ab} * g \) (\kvec ) \equiv \int (d \kvec' ) G_{ab} (\kvec, \kvec') g (\kvec') 
\eeq
The free energy density  then takes the simple form:
\beq
\label{multi.free}
\CF 
= -\inv{\beta} \sum_a  \int (d \kvec )   \[  
-s_a \log \( 1 - s_a e^{-\beta \vep_a } \)  - 
  \frac{  1 -  y_a^{-1} }{2(e^{\beta \vep_a }  - s_a )} \]
\eeq

For the two-component fermion defined by the action
(\ref{fermionaction}),  spin up particles scatter with 
spin down,  and $G_{\up \down} = G_{\down \up}$.   
Thus, if the chemical potentials are equal, 
$\mu_\up = \mu_\down = \mu$,  then $\vep_\up = \vep_\down$,
and the thermodynamics is just two copies of the single component
fermion described above.   In particular, the density and $c$ 
are doubled,  but the energy per particle scaling functions
are the same.

\section{Conclusions}

We have shown that the S-matrix based approach to quantum gases
developed in \cite{PyeTon} leads to a new treatment 
of  the   scale-invariant unitary limit, where all of the
thermodynamic scaling functions can be computed as a function
of $\mu / T$.    Though our methods are of course not exact, 
they are sufficiently novel to provide new insights into these
systems.   It would be worthwhile to undertand the corrections
to our results due to other types of diagrams involving 
N-body scattering for $N>2$,  and also other more complicated
diagrams involving 2-body scattering;  this can be systematically
explored using the full formalism in \cite{PyeTon}.   

In this paper we have mainly analyzed the 2-dimensional case,
deferring the analysis of the integral equations in the 3-dimensional
case to a separate publication\cite{InPrep}.    For the 2-dimensional
case,  this required us to  define a meaningful unitary limit where
the S-matrix equals $-1$, and such a limit has not been considered
before.     We have calculated most of the interesting scaling
functions for the free energy and energy per particle.  

The ratio of the shear viscosity to entropy density $\eta/s$ 
was also analyzed, and for fermions and repulsive bosons,  it 
is above the conjectured lower bound of $\hbar/ 4\pi k_B$.  
For attractive bosons it drops below it by a factor of $0.4$,   however 
this could be an artifact of our approximation.    
For attractive fermions,  the conjectured lower bound is reached 
at very large $\mu/T \approx  10^7$,   however   this  was argued to
occur in an unphysical or metastable  region;  in the physical regions,
$\eta/s \geq  6.07\, \hbar/4 \pi k_B$.

\section{Acknowledgments}

We thank Erich Mueller for helpful discussions.    
   This work is supported by the National Science Foundation
under grant number  NSF-PHY-0757868.

\section{Appendix A:  The beta function}

Consider a single free boson with action eq. (\ref{bosonaction}).  
Setting the external energy and momentum to zero,  the 
4-point vertex at tree-level plus 1-loop is
\beq
\label{A.1}
\Gamma^{(4)} = - i g  -  \frac{(-ig)^2}{2} \int  \frac{d \omega
  d^d \kvec}{(2\pi)^{d+1} }  \(  \frac{i}{w - \kvec^2/2m + i \epsilon}\) 
\( \frac{i}{ -\omega - \kvec^2/2m  + i \epsilon}\) 
\eeq
The $\omega$ integral can be done by closing the contour in 
the upper half plane picking up the pole at
 $\omega = - \kvec^2/2m + i \epsilon$:
\beq
\label{A.2}
{\rm 1-loop} =   - i \frac{ m g^2}{2} \int \frac{d^d \kvec}{(2\pi)^d}  
\inv{\kvec^2} 
\eeq
Introducing an ultraviolet cut-off $\Lambda$:
\beq
\label{A.3}
\Gamma^{(4)}=  - i \(  g -  \frac{m g^2  \Lambda^{d-2}}
{\pi^{d/2} \Gamma( d/2 ) 2^d (d-2)} 
\) 
\eeq
Defining the dimensionless coupling $\ghat$ as 
$g = \Lambda^{2-d} \ghat$, and requiring the $\Gamma^{(4)}$ be
independent of $\Lambda$ gives the beta function:
\beq
\label{A.4}
\frac{d \ghat}{d \ell} =  (2-d) \ghat - \frac{m}{\pi^{d/2} \Gamma(d/2) 2^d}
\ghat^2 
\eeq
where $\ell = - \log \Lambda$ is the logarithm of a length scale;
increasing $\ell$ corresponds to flowing to lower energy.  
With the convention for the coupling $g$ 
 in eq. (\ref{fermionaction}) for the fermionic case,  
the beta-function is the same as above but to half the symmetry
factor of the 1-loop diagram.

\section{Appendix B:  Derivation of the multi-component case}

In this appendix,  we extend the derivation in \cite{PyeTon} to 
the general case of multiple species of particles,  of possibly
mixed Bose/Fermi statistics.  
Following the definitions in section VII,  
  the 2-body foam diagram approximation is obtained by considering
the free energy functional:
\barray
\label{B.1}
\digamma &=&  -\inv{\beta} \int (d \kvec ) \sum_a \(  s_a \log (1 + s_a f_a ) -
\frac{f_a - f_{0,a}}{1 + s_a f_{0,a}} \)  
\\ 
&~& ~~~~~~~~~~
- \inv{2}  \int (d \kvec ) (d\kvec' ) \sum_{a,b} \ftilde_a (\kvec ) 
G_{ab} (\kvec, \kvec' ) \ftilde_b (\kvec' ) 
\earray
where
\beq
\label{B.2}
f_{0,a} =  \inv{e^{\beta(\omega_a - \mu_a)} - s_a } , ~~~~~~~~~
\ftilde_a =  f_a / y_a  
\eeq
The integral equation for the pseudo-energy $\vep_a$ follows from the
variational principle $\delta \Fhat / \delta f_a  = 0$.   Using
\beq
\label{B.3}
\frac{ \delta \ftilde_a (\kvec )}{\delta f_b (\kvec' )} =  \delta_{a,b} 
(2 \pi )^d \delta( \kvec - \kvec' ) \frac{s_a f_{0,a}}{1 + s_a f_{0,a}} ,
\eeq
after some algebra one obtains the integral equation (\ref{multi.integral}).   
(We have used $G_{ab} (k,k') = G_{ba} (k', k)$.)
Using the solution to the integral equation  (\ref{multi.integral}) 
 in the  expression for $\Fhat$,  one
obtains the free energy density $\CF = \Fhat$ in eq. (\ref{multi.free}).

\end{document}